\documentclass[]{emulateapj}
\usepackage{epsf}
\begin{document}

\newcommand{\Vin}{$V_{\rm{max}}^{\rm{acc}}\,$}
\newcommand{\Vmax}{$V_{\rm{max}}\,$}
\newcommand{\Vnow}{$V_{\rm{max}}^{\rm{now}}\,$}

\shorttitle{MODELING GALAXY CLUSTERING THROUGH COSMIC TIME}
\shortauthors{CONROY, WECHSLER \& KRAVTSOV}

\title{Modeling Luminosity-Dependent Galaxy Clustering Through Cosmic Time}

\author{Charlie Conroy, Risa H.  Wechsler\altaffilmark{1}, Andrey V. Kravtsov} 
\affil{Department of Astronomy and Astrophysics, 
  Kavli Institute for Cosmological Physics, \& The Enrico Fermi Institute,\\
  The University of Chicago, 5640 S. Ellis Ave., Chicago, IL 60637}

\altaffiltext{1}{Hubble Fellow}

\begin{abstract}
  
We employ high-resolution dissipationless simulations of the
concordance $\Lambda$CDM cosmology ($\Omega_0=1-\Omega_{\Lambda}=0.3$,
$h=0.7$, $\sigma_8=0.9$) to model the observed luminosity dependence
and evolution of galaxy clustering through most of the age of the
universe, from $z\sim 5$ to $z\sim0$.  We use a simple, non-parametric
model which monotonically relates galaxy luminosities to the maximum
circular velocity of dark matter halos ({\Vmax}) by preserving the
observed galaxy luminosity function in order to match the halos in simulations
with observed galaxies.  The novel feature of the model is the use of
the maximum circular velocity at the time of accretion, {\Vin}, for
subhalos, the halos located within virial regions of larger halos.  We
argue that for subhalos in dissipationless simulations, {\Vin}
reflects the luminosity and stellar mass of the associated galaxies
better than the circular velocity at the epoch of observation,
{\Vnow}.  The simulations and our model $L-${\Vmax} relation predict
the shape, amplitude, and luminosity dependence of the two-point
correlation function in excellent agreement with the observed galaxy
clustering in the SDSS data at $z\sim 0$ and in the DEEP2 samples at
$z\sim 1$ over the entire probed range of projected separations,
$0.1<r_p/(h^{-1}$ Mpc$)<10.0$.  In particular, the small-scale upturn
of the correlation function from the power-law form in the SDSS and
DEEP2 luminosity-selected samples is reproduced very well. At $z\sim
3-5$, our predictions also match the observed shape and amplitude of
the angular two-point correlation function of Lyman-break galaxies
(LBGs) on both large and small scales, including the small-scale
upturn.  This suggests that, like galaxies in lower redshift samples,
the LBGs are fair tracers of the overall halo population and that
their luminosity is tightly correlated with the circular velocity (and
hence mass) of their dark matter halos.

\end{abstract}


\section{Introduction}\label{s:intro}

A generic prediction of high-resolution simulations of hierarchical
Cold Dark Matter (CDM) models is that virialized regions of halos are
not smooth, but contain subhalos --- the bound, self-gravitating dark
matter clumps orbiting in the potential of their host halo. The
subhalos are the descendants of halos accreted by a given system
throughout its evolution, which retain their identity in the face of
disruption processes such as tidal heating and dynamical
friction. Their presence is in itself a vivid manifestation of the
hierarchical build-up of halo mass.

In the CDM scenario, luminous galaxies form via cooling and
condensation of baryons in the centers of the potential wells of dark
matter halos \citep{White78,Fall80,Blumenthal84}. In the context of
galaxy formation, there is little conceptual difference between halos
and subhalos, because the latter have also been genuine halos and
sites of galaxy formation in the past, before their accretion onto a
larger halo. We thus expect that each subhalo of sufficiently large
mass should host a luminous galaxy and this is indeed supported by
self-consistent cosmological simulations \citep[e.g.,][]{Nagai05}.
The observational counterparts of subhalos are then galaxies in
clusters and groups or the satellites around individual galaxies.  In
this sense, we will use the term {\it halos} to refer to both distinct
halos (i.e., halos not located within the virial radius of a larger
system) and subhalos.

Although this general picture is definitely reasonable, it is not
clear just how direct the relation between halos and galaxies is.  One
may argue, for example, that the subhalos can be disrupted much faster
than the more tightly bound stellar system they host, leaving behind
``orphan'' galaxies (\citealt{Gao04a}; \citealt*{Diemand04}). At the
same time, properties of surviving subhalos, such as maximum circular
velocity and gravitationally bound mass, are subject to strong
dynamical evolution as they orbit within the potential of their host
halo \citep[e.g.,][]{Moore96,Klypin99,Hayashi03,Kravtsov04b,Kazantzidis04b}. This
makes the relation between subhalo properties and galaxy luminosity
ambiguous \citep{Nagai05}, because the latter may be less affected by
dynamical processes but may evolve due to aging of the stellar
populations after ram pressure strips the existing gas and the
accretion of new gas is suppressed.  The key question that we address
in this paper is whether there is a one-to-one correspondence between
populations of halos in dissipationless cosmological simulations and
galaxies in the observable universe. As a test, we use comparisons of
the predicted clustering of halos with the available observational
measurements of galaxy clustering from $z\sim 5$ to the present.

During the last decade, large observational surveys of galaxies both at
low and high redshifts have tremendously improved our knowledge of
galaxy clustering, its evolution, and the relation between the galaxy and
matter distributions.  A coherent picture has emerged in which bright
galaxies are strongly biased with respect to the matter distribution at
high redshifts \citep{Steidel98,Giavalisco98,Adelberger03,
  Adelberger05, Ouchi04b, Ouchi05, Lee05, Hamana05}, and in which the bias 
decreases with time in such a way that the amplitude of galaxy
clustering is only weakly evolving \citep[e.g.,][]{Ouchi04b}, as
expected in hierarchical structure formation \citep{Colin99,
  Kauffmann99}.  The bias is also in general scale- and
luminosity-dependent.  Bright (red) galaxies are more strongly
clustered than faint (blue) galaxies both in the local universe
\citep[][and references therein]{Norberg02a, Zehavi04, Zehavi05} and
in the distant past \citep{Coil04a, Coil05a}.

A recent development is the detection of a departure from a pure power
law in the two-point correlation function of galaxies at $z\sim 0$
\citep{Zehavi04, Zehavi05}. This departure is expected in general
because the two-point correlation function is a sum of two separate
contributions: the one-halo term, which arises from pairs of galaxies
within a distinct dark matter halo, and the two-halo term, which
arises from pairs of galaxies from two different distinct halos
\citep[e.g.,][]{Cooray02}. The one-halo contribution dominates on
small scales, while at scales larger than the size of the largest
virialized regions clustering is due to the two-halo term. The two
terms are not generically expected to combine so as to give a power-law
correlation function. The deviation of the correlation function from a
power-law was {\it predicted} to be even stronger at higher redshifts
\citep{Zheng04,Kravtsov04}, and this has now been convincingly confirmed
\citep{Adelberger05, Ouchi05, Lee05, Hamana05} using galaxy samples
identified with the Lyman-break technique \citep{Steidel96,Steidel99}.

Despite impressive advances in the amount and quality of data on the
galaxy distribution over a wide range of redshifts, the exact relation
between dark matter halos and luminosity- or Lyman-break-selected
galaxies is still rather uncertain \citep[e.g.,][]{Mo99,Kolatt99,
Wechsler01}.  The most popular attempts to connect the dark
halos and luminous galaxies employ semi-analytic modeling
(e.g., \citealt{White91,Kauffmann94}; \citealt{AvilaReese98,
Somerville99,Cole94,Cole00, Croton05,Bower05}), which uses 
phenomenological recipes for specifying when, where, and how galaxies
form within dark matter halos, often in conjunction with
high-resolution dissipationless simulations. Another popular approach
is to use the halo model \citep[see e.g.][and references
therein]{Cooray02}, which, in its simplest form, specifies the
probability distribution for a particular halo of mass $M$ to host a
given number of galaxies $N$ with specified properties, such as
luminosity, color, etc.  More complex halo models include the
conditional luminosity function (CLF) approach
\citep{Yang03,Cooray05a, Cooray05b} which specifies the luminosity
function for halos of mass $M$, and models which connect other features
\citep*[e.g.,][]{Vale04, Neyrinck05}.

While these approaches manage to capture the general observational
trends, they usually employ a large number of free parameters, making
it difficult to glean the relevant underlying mechanism(s) responsible
for the agreement between the model and data.  Most implementations
also do not provide a direct constraint on the relation between
luminous components of galaxies and dark matter halos. In fact, {\it
  assumptions} are made about this relation in most cases.

In this study we use galaxy clustering to address two straightforward
issues.  First, we question whether current data is
consistent with a one-to-one correspondence between
luminous galaxies and galaxy-sized dark matter halos in cosmological
simulations. Second, we ask whether the observed clustering is consistent
with a simple relation between galaxy luminosity and some property of
its host halo.  In the context of the second question, we investigate
which halo properties are most closely related to galaxy luminosity.
We show with the currently available data the answer to both of these
questions is yes, as we can reproduce luminosity-dependent clustering
measurements at different redshifts from $z\sim 5$ to $z\sim 0$
with a simple, non-parametric model relating galaxy luminosity
to the halo circular velocity.

Although a number of studies during the last decade have shown that
galaxy clustering can be approximately matched by the clustering of
dark matter halos in dissipationless simulations (\citealt{Carlberg91,
  Brainerd92, Brainerd94a, Brainerd94b}; \citealt*{Colin97};
\citealt{Wechsler98, Colin99, Kravtsov99b,
  Kravtsov04,Neyrinck04}) the size of observational samples did
not allow thorough tests of the galaxy-halo relation. For example,
\citet{Kravtsov04} and \citet{Neyrinck04} compare the two-point
correlation function of halos and bright galaxies from the SDSS and
PSCz surveys, respectively, and find good agreement on scales
$0.1\lesssim r/$($h^{-1}$~Mpc$) \lesssim 10$.  However, these studies
do not attempt to match the clustering of fainter galaxies.
\citet{Tasitsiomi04} assign luminosities to halos and compute the
galaxy-mass correlation function, finding good agreement for two broad
luminosity bins, after a reasonable amount of scatter was introduced
into the relation.

The current work extends these analyses by comparing results of very
high-resolution simulations to the most current measurements of the
two-point correlation functions over a wide range of luminosities and
redshifts. The large size and wide luminosity range of observational
samples allows us to test the relation between galaxy luminosities and
properties of their host halos with unprecedented power.  The novel
feature of the simulation analysis we present is that for each halo
and subhalo we track the evolution of its properties, such as mass and
maximum circular velocities. As we show below, this is a key piece of
information for reasons that are easy to understand. For distinct
halos, the current circular velocity is a measure of their potential
well built-up during evolution, and can therefore be expected to be
tightly correlated with the stellar mass (or more generally the
baryonic mass) of the galaxy the halo hosts. The circular velocity of
subhalos in dissipationless simulations, on the other hand, is a
product of both mass buildup during the period when the halo evolved
in isolation {\it and} tidal mass loss after the halo starts to orbit
within the virialized region of a larger object and experience strong
tidal forces \citep{Hayashi03,Kravtsov04b,Kazantzidis04b}. The stellar
component of the galaxies, which should be more tightly bound than
halo dark matter, should not be significantly affected by tidal forces
and can stabilize the mass distribution (and hence {\Vmax}) in the
inner regions. We can therefore expect that luminosity and stellar
mass of galaxies hosted by halos in dissipationless simulations should
be correlated with the subhalo mass or circular velocity, \Vin, {\it
  at the epoch of accretion}, rather than with its current value.
This is borne out by cosmological simulations, which include gas
dynamics, cooling, and star formation \citep{Nagai05} who show that
selection using {\Vin} results in subhalo properties similar to the
selection based on stellar mass of galaxies subhalos host.  One can
therefore argue that a reasonable approach is to relate galaxy
luminosity to the current halo circular velocity for distinct halos
and to the circular velocity at accretion for subhalos. The main
result of this study is that this simple model reproduces the
luminosity-dependence of galaxy clustering at different epochs with
remarkable, and perhaps surprising, accuracy.  We note that
\citet{Vale05} have recently presented and used a semi-analytic model
for subhalos, which employs a similar approach to luminosity
assignment, except that they use the total bound halo mass instead of
circular velocity.

The paper is organized as follows.  In $\S$\ref{s:sims} we briefly
describe the simulations, halo finding algorithm, and the method for
tracking the evolution of halos.  $\S$\ref{s:appr} details our method for
relating halos to galaxies, and elaborates upon our motivation for
using \Vin as the basis for the luminosity assignment for subhalos.  In
$\S$\ref{s:res} we compare observational clustering results over the
redshift interval $0<z<5$ to the clustering of halos in
dissipationless $\Lambda$CDM simulations.  The halo occupation
distribution implied by this model is described in $\S$\ref{s:hod}.
In $\S$\ref{s:disc} we discuss the implications of our results and in
$\S$\ref{s:conc} we summarize our main conclusions.  Throughout this
paper we assume a $\Lambda$CDM cosmology with
$(\Omega_m,\Omega_{\Lambda},h,\sigma_8) = (0.3,0.7,0.7,0.9)$.

\section{The Simulations}\label{s:sims}

The simulations used here were run using the Adaptive Refinement Tree
(ART) $N$-body code \citep{Kravtsov97, Kravtsov99a}.  The ART code
implements successive refinements in both the spatial grid and
temporal step in high density environments.  These simulations were
run in the concordance flat $\Lambda$CDM cosmology with
$\Omega_m=0.3=1-\Omega_{\Lambda}$, $h=0.7$, where $\Omega_m$ and
$\Omega_{\Lambda}$ are the present-day matter and vacuum densities in
units of the critical density, and $h$ is the Hubble parameter in
units of $100$ km s$^{-1}$ Mpc$^{-1}$.  The power spectra used to
generate the initial conditions for the simulations were determined
from a direct Boltzmann code calculation (courtesy of Wayne Hu).  We
use a power spectrum normalization of $\sigma_8=0.90$, where
$\sigma_8$ is the rms fluctuation in spheres of $8h^{-1}$ Mpc comoving
radius. 

To study the clustering properties of dark matter halos over a range
of scales, we consider two simulations of the above cosmology
The first simulation, L80,
followed the evolution of $512^3$ particles in a $80h^{-1}$ Mpc box.
The L80 simulation has a particle mass of $m_p=3.16\times 10^8$
$h^{-1} M_\Sun$ and peak force resolution $h_{\rm{peak}}=1.2$ $h^{-1}$
kpc.  This is the simulation to which we make most comparisons with
observations.  The second simulation we consider is denoted L120 and
was run with $512^3$ particles in a $120$ $h^{-1}$ Mpc box, resulting
in a particle mass of $m_p=1.07\times 10^9$ $h^{-1} M_\Sun$ and peak
force resolution $h_{\rm{peak}}=1.8$ $h^{-1}$ kpc.  This simulation
thus has a larger particle mass and somewhat lower spatial resolution
compared to the L80 run.  We use this simulation to obtain better
statistics for the correlation function of rare (i.e., massive)
objects.

\begin{figure}
\plotone{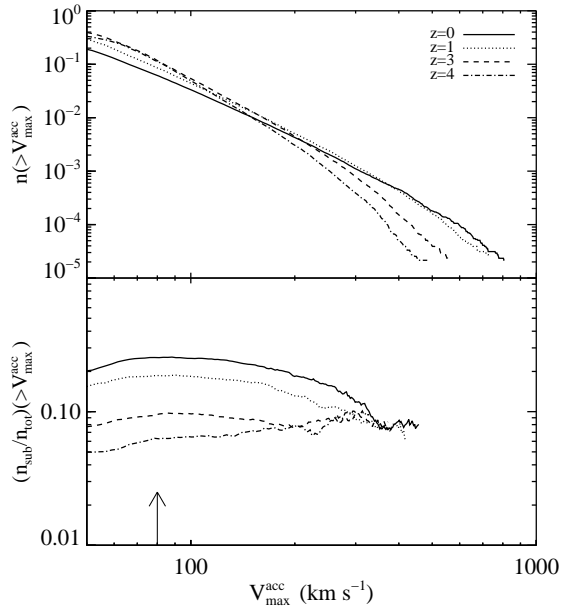}
\vspace{0.5cm}
\caption{{\it Top panel}: Cumulative velocity function for all halos
identified in the L80 simulation at various redshifts, in units of
$h^3$ Mpc$^{-3}$.  {\it Bottom panel}: Fraction of subhalos as a
function of redshift and maximum circular velocity at the time of
accretion, \Vin.  We truncate the curves where $N_{\rm{sub}}<10$
because in that regime poisson noise washes away any useful
information.  The arrow delimits our nominal completeness limit.}
\label{f:cumn_z}
\end{figure}

\subsection{Halo Identification, Classification, and Construction of Merger Trees}

Our analysis requires detailed dynamical knowledge of not only
distinct halos, i.e. halos with centers that do not lie within any
larger virialized system, but also subhalos which are located within
the virial radii of larger systems.  Note that the term ``halo''
(e.g., the {\it halo} occupation distribution) usually refers to what
we call distinct halos in this work.

We identify distinct halos and the subhalos within them using a
variant of the Bound Density Maxima (BDM) halo finding algorithm
\citep{Klypin99}.  Details of the algorithm and parameters used can be
found in \citet{Kravtsov04}; we briefly summarize the main steps here.
All particles are assigned a density using the \texttt{smooth}
algorithm\footnote{To calculate the density we use the publicly
available code \texttt{smooth:
http://www-hpcc.astro.washington.edu/tools/ tools.html}} which uses a
symmetric SPH smoothing kernel on the $32$ nearest neighbors.
Starting with the highest overdensity particle, we surround each
potential center by a sphere of radius $r_{\rm{find}}=50h^{-1}$ kpc
and exclude all particles within this sphere from further search.
Hence no two halos can be separated by less than $r_{\rm{find}}$.  We
then construct density, circular velocity, and velocity dispersion
profiles around each center, iteratively removing unbound particles as
described in \citet{Klypin99}.  Once unbound particles have been
removed, we measure quantities such as $V_{\rm{max}} =
\sqrt{GM(<r)/r}|_{\rm{max}}$, the maximum circular velocity of the
halo.  For each distinct halo we calculate the virial radius, defined
as the radius enclosing overdensity of 180 with respect to the {\it
mean} density of the Universe at the epoch of the output. We use this
virial radius to classify objects into distinct halos and subhalos.
The halo catalog is complete for halos with more than $50$ particles
which corresponds, for the L80 simulations, to a minimum halo mass of
$1.6\times 10^{10} h^{-1}$ $M_{\Sun}$.

For subhalos we also tabulate \Vin, the maximum circular velocity at
the time when a subhalo falls into a distinct halo.  In order to
tabulate this quantity we rely on merger trees 
generated for these simulations, which track the histories of both
distinct halos and subhalos.  A detailed description of the merger
tree construction is given in \citet{Allgood05}.  The merger trees we
use here follow halo evolution through 48 timesteps between $z=7$ and
the present for the L80 box and 89 timesteps for the L120 box.  For
each subhalo, we use the merger trees to step back in time until the
subhalo is no longer identified as belonging to a larger system.  We
then define \Vin to be \Vmax of the subhalo at that time.

In the top panel of Figure \ref{f:cumn_z} we show the cumulative
velocity function for all identified halos from $z=4$ to $0$.  This
figure quantifies the relation between \Vin and $n$, as we will use
these two quantities interchangeably to define our halo samples
throughout the paper. In the bottom panel we show the corresponding
cumulative fraction of subhalos as a function of time. The figure
shows that the subhalo fraction is a weak function of circular
velocity at all epochs. There is a weak trend for a smaller subhalo
fraction among halos with larger \Vin. There is a stronger trend of
increasing subhalo fraction with decreasing redshift.

\section{Connecting Galaxies to Halos}\label{s:appr}

In this section we motivate and describe our model for associating
galaxies with dark matter halos.  We make this connection by relating
galaxy luminosity and a physical property of dark matter halos, for
which we choose \Vnow for distinct halos and \Vin for subhalos.  As
discussed above, for subhalos \Vin denotes the maximum circular
velocity \emph{at the time the subhalo was accreted}.  The maximum of the
circular velocity profile, \Vmax, is a measure of the depth of the
halo potential well and is expected to correlate strongly with stellar
mass of the galaxies, as implied by the Tully-Fisher and Faber-Jackson
relations. At the same time, the definition of \Vmax in simulations is
unambiguous both for distinct halos and subhalos, which is not the
case for the total mass, as different operational definitions are used
by different authors.  It should be noted that \Vmax used here will
not correspond directly to \Vmax observed in, for example, the
Tully-Fisher relation because dissipationless simulations do not take
into account the effect of baryon condensation on \Vmax
\citep[e.g.,][]{Blumenthal86}.

\begin{figure}
\plotone{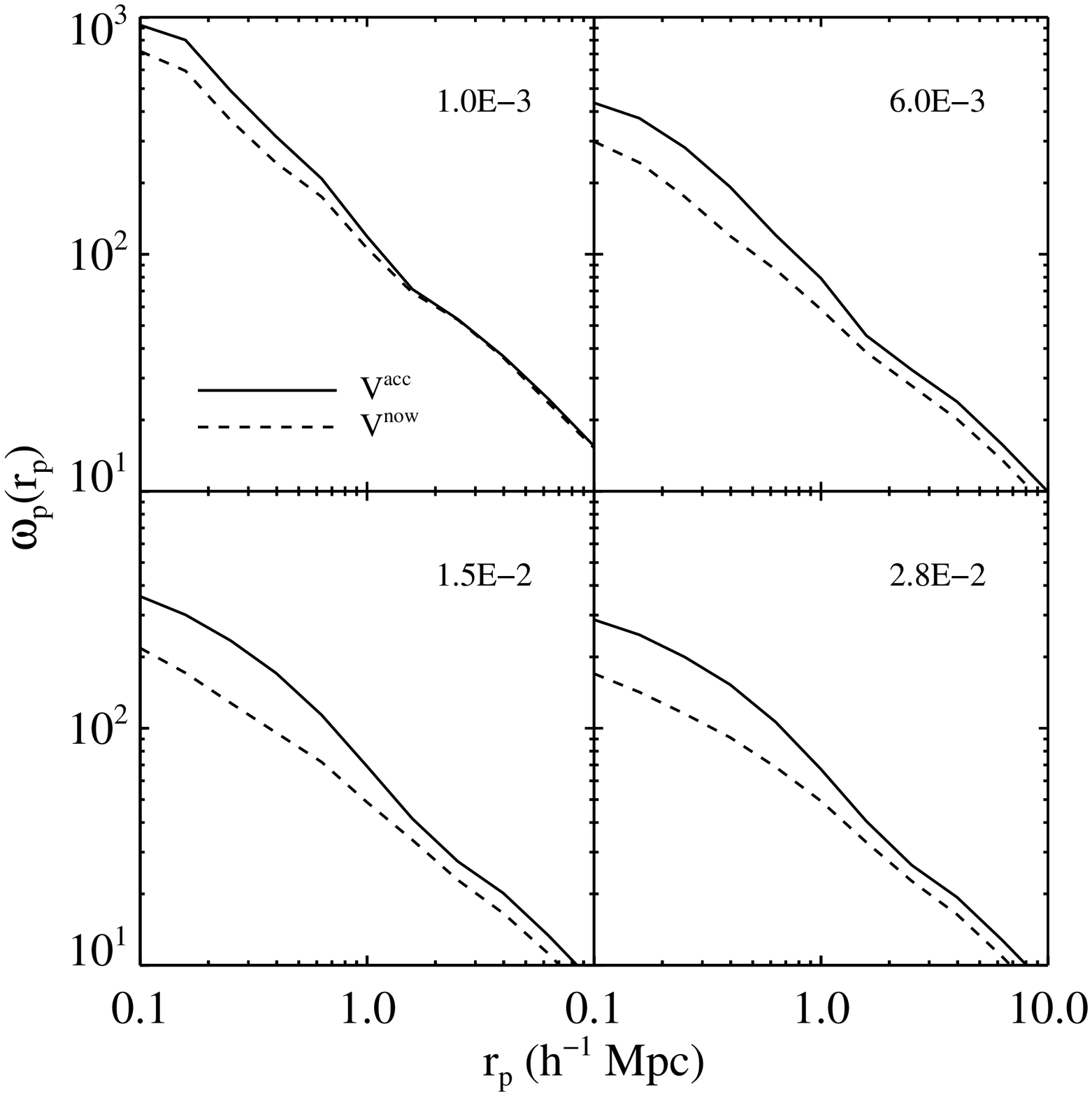}
\vspace{0.5cm}
\caption{Projected two-point correlation function at $z\sim0$
comparing the effects of selecting on \Vin (solid lines) versus \Vnow
(dashed lines) at four different number density thresholds (labeled in
the top right corner, in units of $h^3$ Mpc$^{-3}$).  While there is a
slight increase in the correlation function on large scales when using
\Vin rather than \Vnow, the difference is much stronger on small
scales.  The difference between \Vin and \Vnow is due to the tidal
stripping of subhalos which have fallen into larger systems, hence
correlation functions will be most strongly effected on small scales.}
\label{f:vcomp}
\end{figure}

The use of \Vin for subhalos is the novel feature of our
model\footnote{As we prepared this paper for publication,
\citet{Vale05} submitted a paper in which they employ a semi-analytic
model for subhalos and a similar approach to luminosity assignment,
except that they use the total bound halo mass instead of circular
velocity.}. As we discussed in the introduction, the motivation for
this is fairly straightforward. While \Vmax decreases due to tidal
stripping as a halo falls through a larger halo
\citep{Hayashi03,Kravtsov04b}, one can expect that the stellar component of
galaxies within these halos will not be affected appreciably since
stars are concentrated near the bottom of the halo potential well and
are more tightly bound \citep[e.g.,][]{Nagai05}.  Hence we argue that,
for galaxies associated with subhalos, luminosity should correlate
more strongly with {\Vin} than the {\Vnow} affected by dynamical
evolution.  Therefore, throughout the rest of the paper, unless
explicitly stated otherwise, the maximum circular velocity, \Vmax,
will be assumed to mean:
\begin{equation}
V_{\rm max} =\left\{ \begin{array}{ll} V_{\rm max}^{\rm acc}, & {\rm
  subhalos}\\[1mm] V_{\rm max}^{\rm now}, & {\rm distinct\ halos}
\end{array}
\right.
\end{equation}
Although it is not clear how accurate this assumption is in detail, we
show that it provides a considerably better match to observed galaxy
clustering compared to the uniform selection using {\Vnow} for both
subhalos and distinct halos. Note also that the use of circular
velocities before accretion can also help explain the abundance of
faint dwarf galaxies in the Local Group \citep{Kravtsov04b}.

In order to assign luminosities, we assume a monotonic relation
between galaxy luminosity and {\Vmax} and require that the $L-${\Vmax}
relation preserves the galaxy luminosity function (LF).  Specifically,
we use the following equation:
\begin{equation}\label{eqn:v2l}
n_g(>L_i) = n_h(>{V_{\rm{max}}}_{,i})
\end{equation}
where $n_g$ and $n_h$ are the number density of galaxies and halos,
respectively. We stress again, that {\Vmax} in the above expression is
equal to {\Vin} for subhalos and to {\Vnow} for distinct halos (for
which the accretion epoch is undefined).  For each $L_i$ we find the
corresponding $V_{\rm{max},i}$ such that the above relation is
satisfied.  The main assumption is therefore that there is a monotonic
relation between galaxy luminosity and {\Vmax}. The model makes no
further assumptions and is completely non-parametric. This should be
kept in mind when we compare {\it predictions} of this model to
observed galaxy clustering.  Note that we do not take any possible
scatter in the $L-${\Vmax} relation into account in this model.

With the $L-${\Vmax} relation in hand, comparing observational
clustering statistics to the model predictions is straightforward: we
simply compute the desired statistic for the halos with assigned
luminosities corresponding to the observed sample luminosity range.
This method does not currently treat other galaxy properties such as
color, although it could conceivably be extended to include such
properties. We have not included the possibility of ``orphaned
galaxies'', i.e. galaxies without any associated subhalos.  We discuss
the issue of orphans in detail in $\S$\ref{s:disc}.

Figure \ref{f:vcomp} shows the effect that selection of halos using
\Vin rather than \Vnow for subhalos has on the projected two-point
correlation function $\omega_{\rm p}$ for different number densities,
at $z=0$.  As expected, the effect is most significant on small
scales, where the subhalo contribution dominates, and  the difference
between \Vin and \Vnow is greatest.

\begin{figure}
\plotone{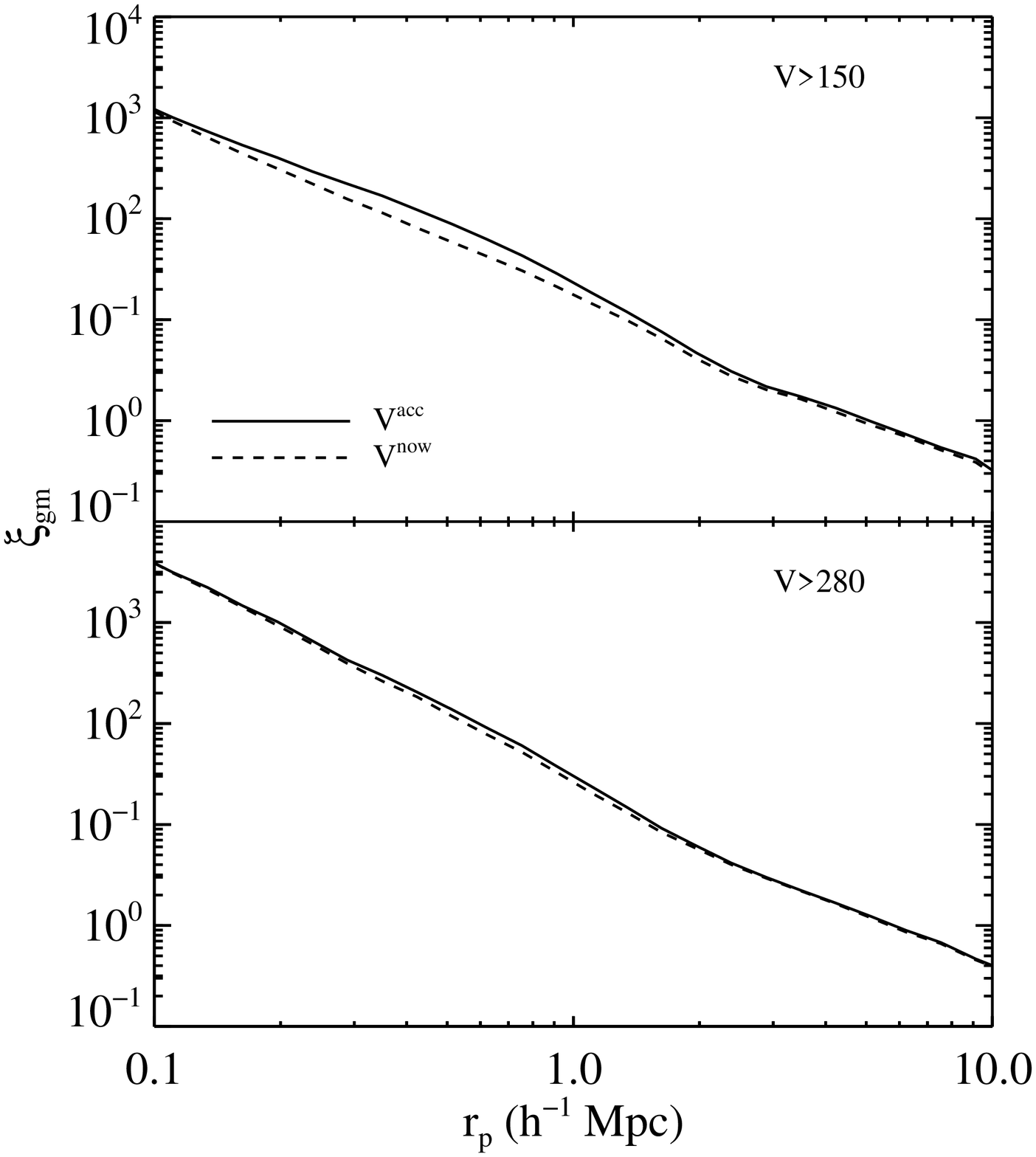}
\vspace{0.5cm}
\caption{Comparison of the galaxy-mass cross-correlation function for
halos selected using \Vin (solid lines) and \Vnow (dashed lines) at
two different circular velocity thresholds (labeled in the top right
corner of each panel, in units of km s$^{-1}$).}
\label{f:xigm}
\end{figure}

In Figure \ref{f:xigm} we show the effect of selecting halos according
to \Vin and \Vnow on the galaxy-mass cross-correlation function,
$\xi_{gm}$.  The small bump in the sample selected using \Vin in the
top panel is due to the fact that \Vin samples in general have more
satellites than \Vnow samples, and the satellite contribution to
$\xi_{gm}$ peaks near $r_{\rm p}=0.5$ $h^{-1}$ Mpc.  The ``bump'' is
smaller in the lower panel, where the \Vmax threshold is much higher,
because the number of satellites is nearly the same between the \Vin
and \Vnow selected samples in this case.

\begin{figure}
\plotone{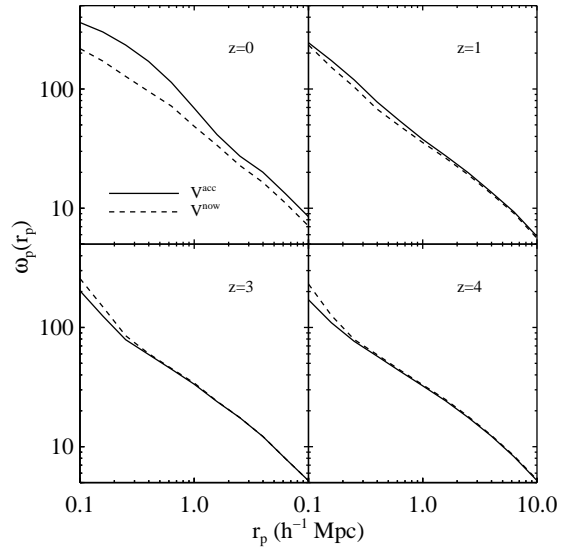}
\vspace{0.5cm}
\caption{Comparison of the projected two-point correlation function
for halos selected using \Vin (solid lines) and \Vnow (dashed lines)
at four different redshifts, for a fixed number density,
$n=1.5\times10^{-2} h^3$ Mpc$^{-3}$.  This figure clearly shows that,
while using \Vin over \Vnow results in a large difference at low
redshift, it has very little impact at higher redshifts.  The trend is
similar for a wide range of number densities.}
\label{f:vin_ev}
\end{figure}

At higher redshifts the differences between samples selected using
{\Vin} and {\Vnow} are small.\footnote{Note that the designation
``now'' refers to the time of observation, not to $z=0$.}  Figure
\ref{f:vin_ev} shows the projected correlation function for the \Vin-
and \Vnow-selected samples at four redshifts. The samples are
constructed to have a fixed number density $n=1.5\times10^{-2} h^3$
Mpc$^3$.  The situation is similar for other number densities.
Already by $z\sim1$ the effect of selection is quite small.  The
difference on small scales for the $z\sim3$ and $z\sim4$ samples is
not statistically significant.\footnote{Note that at high $z$ the
correlation function at the smallest scales is somewhat higher for the
\Vnow-selected samples, which seems counter-intuitive.  We believe
that this is a small artefact of the halo finding algorithm. At higher
redshifts, the halos are smaller and subhalos are typically at smaller
distances from the host center.  At small radii the removal of unbound
particles is more difficult as the halos are located near the bottom
of the potential well. The value of \Vnow in such cases can be
overestimated producing a boost in the number of subhalos above a
given threshold value, and boosting the correlation function somewhat.
Note, however, that the effect is small and the difference between
correlation functions for \Vnow- and \Vin-selected samples is less than $2\sigma$.}

We believe that this effect can be attributed to the distribution of
accretion times for subhalos at each redshift: at low redshift,
subhalos have a wide distribution of accretion times, and hence a
large number of subhalos have had time to experience significant tidal
stripping, while at higher redshifts the distribution of accretion
times rises sharply near the epoch of observations. This is because
both the accretion and disruption rates are high at high
redshifts. The accreted halos do not survive for a prolonged period of
time, so that at each high-$z$ epoch most of identified subhalos are
recently accreted objects, which are yet to experience significant
tidal mass loss.

Since we are only computing \Vin for subhalos which have survived to
the current epoch, one may worry that we are neglecting a significant
population of subhalos with a sizeable \Vin that are not present in
the halo catalog at the current epoch.  Of such a population there can
only be two fates: either the object was at some point physically
disrupted or it has simply fallen below the resolution limit.  We have
used the merger trees to find all subhalos which have ever fallen into
a distinct halo and have tabulated their \Vin and the ratio in mass
between the subhalo and distinct halo, at the time of accretion.  For
a wide range in \Vin thresholds, the distribution of mass ratios is
strongly peaked between $0.1$ and $1.0$.  This implies that dynamical
friction has caused the subhalo to merge with the distinct halo on the
order of a dynamical time, and suggests that the majority of these
subhalos have in fact physically merged and should not have survived.
The absence of such missing subhalos also implies that in our
simulations there should be no ``orphan'' galaxies \citep[][see
\S~\ref{s:disc} for further discussion of this issue]{Gao04a}.

\section{Galaxy Clustering From $z\sim5$ to the Present}\label{s:res}

\begin{figure*}
\plottwo{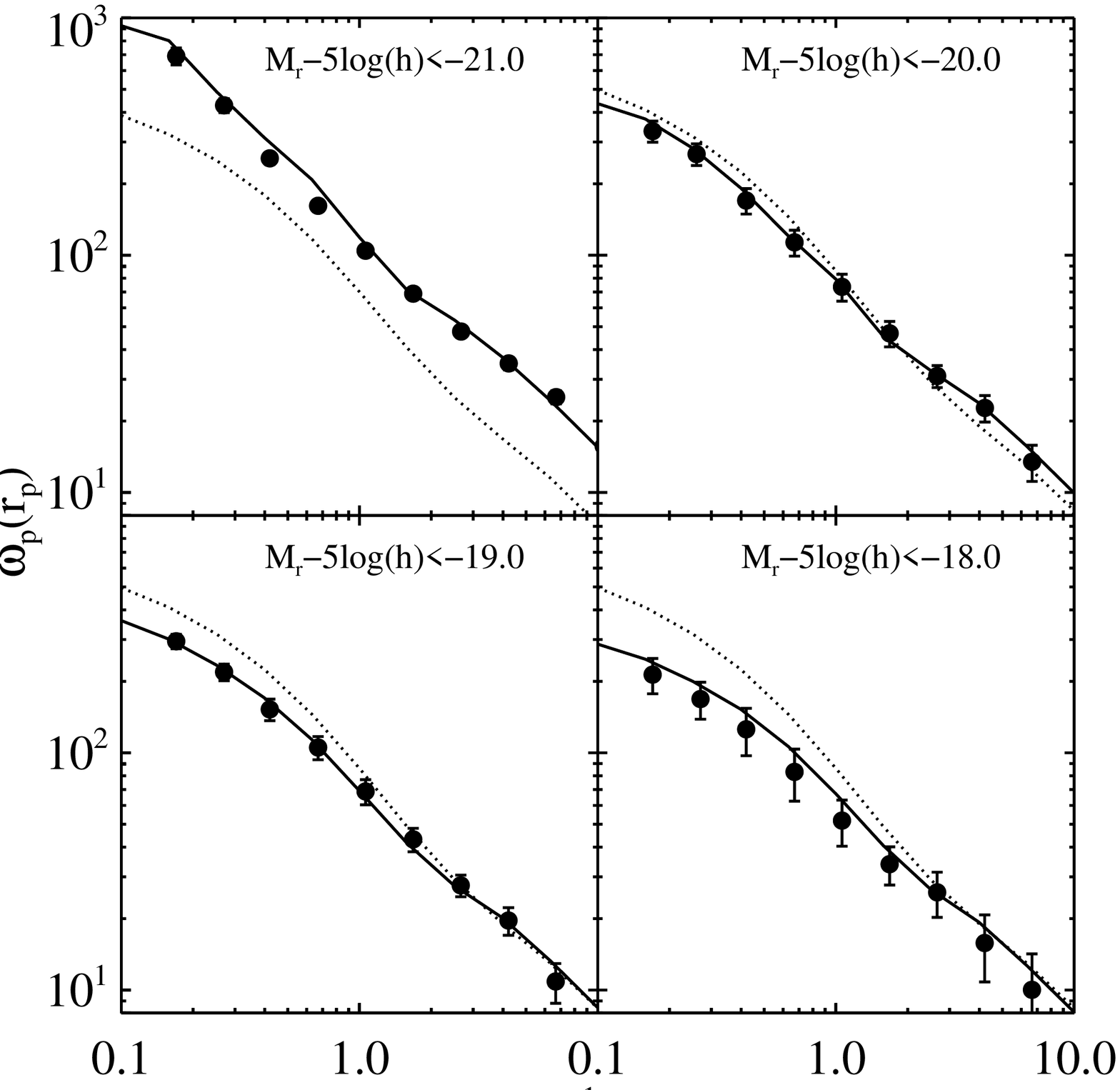}{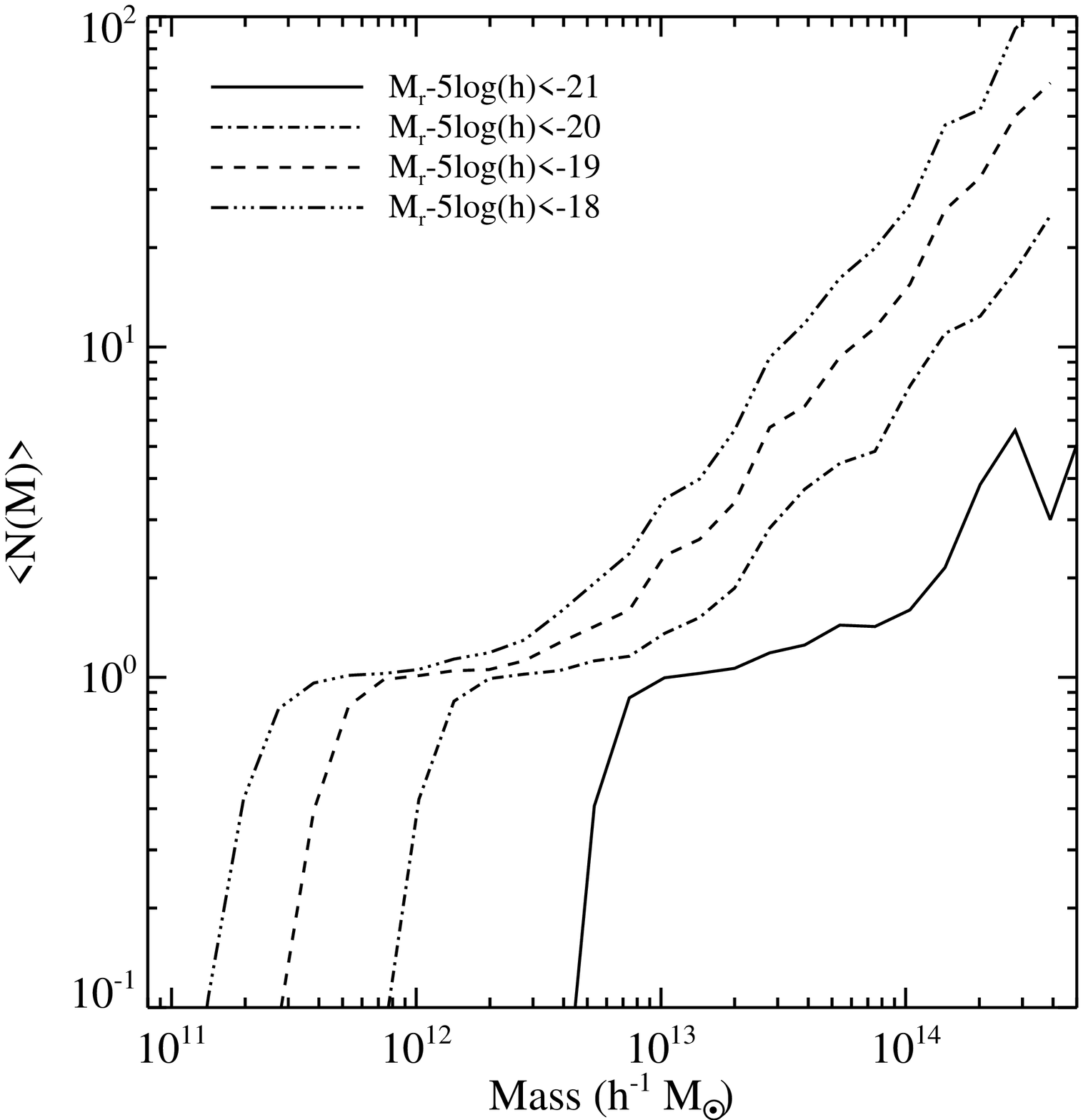}
\vspace{0.5cm}
\caption{{\it Left:} Comparison between the SDSS projected correlation
  function (points) and the correlation function derived from halos
  (solid lines) for various luminosity threshold samples. For
  comparison we include the correlation function of dark matter
  particles (dotted lines) at the median redshift of the sample.  {\it
  Right:} The first moment of the halo occupation distribution (HOD)
  for the four halo samples.  For all four samples, the gradual
  roll-off at small mass is due to scatter in the \Vmax-mass relation.
  The fan (dotted lines) corresponds to slopes of 0.4, 0.7, and 1.0.}
\label{f:sdss_xi}
\end{figure*}

In this section we compare clustering statistics of halos to recent
observations of the galaxy two-point correlation function over the
redshift interval $0\lesssim z\lesssim 5$.  The observed clustering
statistics we compare to are $\omega_p(r_p)$, the projected two-point
correlation function, and $\omega(\theta)$, the angular two-point
correlation function.   We estimate $\omega_p$ by integrating the real
space, three-dimensional correlation function,  $\xi(r)$, computed in
the simulation, along the line of sight:
\begin{equation}\label{eqn:wp}
\omega_p(r_p) = 2\int_0^\infty \xi(\sqrt{r_p^2+r_\parallel^2})
dr_\parallel,
\end{equation}
where the comoving distance $r$ has been decomposed into perpendicular
($r_p$) and parallel to the line-of-sight ($r_\parallel$) components.
In practice, the integration in Eqn.~\ref{eqn:wp} is truncated at some
finite scale: we truncate at $40 h^{-1}$ Mpc for SDSS galaxies
($\S$\ref{s:sdss}) and $20 h^{-1}$ Mpc for DEEP2 galaxies
($\S$\ref{s:deep2}), as is done in the data.  Since the simulation box
size is only $80 h^{-1}$ Mpc, the measurement of $\xi(r)$ is not
reliable for $r\gtrsim 0.1L_{\rm box}\sim 8 h^{-1}$ Mpc. To
extrapolate $\xi(r)$ to larger scales, we use $\xi_{m}(r)$, the
two-point correlation function of dark matter
\footnote{We derive the dark matter correlation function from the
 power spectrum provided by the publicly available code of
 \citet{Smith03}, which is more accurate than the popular Peacock and
 Dodds prescription.}, multiplied by the linear bias of $\xi(r)$
 measured over $4<r/(h^{-1}$Mpc)$<8$.

Generating $\omega(\theta)$ from $\xi(r)$ without assuming $\xi(r)$ to
be a power law is somewhat more involved.  With knowledge of the
redshift distribution, $N(z)$, of the sample, $\omega(\theta)$ can be
derived via the Limber transformation:
\begin{equation}
\omega(\theta) = \frac{\int_0^\infty dz N^2(z) \int^\infty_{-\infty}
dx \xi(\sqrt{[D_m(z)\theta]^2 + x^2})/R_H(z)}{[\int_0^\infty dzN(z)]^2}
\end{equation}
where $D_m(z)$ is the proper motion distance and $R_H(z)$ is the
Hubble radius at redshift $z$.  As for $\omega_{\rm p}$, the integral
over $\xi(r)$ is in practice truncated at a finite scale; we integrate
to $50$ $h^{-1}$ Mpc and note that the resulting $\omega(\theta)$ is
not sensitive to this particular truncation scale.

\subsection{Clustering at $z\sim0$}\label{s:sdss}

The SDSS \citep{York00,Abazajian04} is a large photometric and
spectroscopic survey of the local Universe.  \citet{Zehavi05} have
measured the luminosity and color dependence of $w_p(r_p)$ for
$\sim200,000$ SDSS galaxies over $\approx 2500$ deg$^2$ with $z<0.15$.
As mentioned in $\S$\ref{s:appr}, assigning \Vmax to galaxy luminosity
while preserving the observed luminosity function (LF) results in a
unique $L-${\Vmax} relation.  In order to make the assignment we use
the SDSS luminosity function presented in \citet{Blanton03c}, with
\citet{Schechter76} parameters in the $r$-band
$M^\ast_r-5{\rm{log}}h=-20.5$, $\alpha=-1.05$, and
$\phi^\ast=1.5\times10^{-2} h^{-3}$ Mpc$^3$.  It is then
straightforward to compare the observed luminosity dependence of both
the small and large scale clustering of SDSS galaxies to our model.

The results for luminosity threshold samples ($L>L_{\rm th}$) are
shown in the left panel of Figure \ref{f:sdss_xi}, where we compare
the \citet{Zehavi05} results to the clustering of halos corresponding
to the range of galaxy luminosities in each sample.  For the three
halo samples with $n=6\times10^{-3}h^3$ Mpc$^{-3}$,
$n=1.5\times10^{-2}h^3$ Mpc$^{-3}$ and $n=2.8\times10^{-2}h^3$
Mpc$^{-3}$ we use the L80 simulation, while for the halo sample with
$n=1.1\times10^{-3}h^3$ Mpc$^{-3}$ we use the L120 simulation in order
to improve statistics.  See Table~\ref{t:summary} for details of the
SDSS samples used here.  The agreement is excellent over all scales.
We find similar agreement when $w_p(r_p)$ is measured in differential,
rather than integral, luminosity bins.  It is critical to realize that
the agreement on scales $r_p\lesssim 1 h^{-1}$ Mpc is due to our
luminosity assignment scheme using {\Vin}. The luminosity assigned
using {\Vnow} for subhalos would result in a significant
under-prediction of amplitude of $\omega_{\rm p}$ at $r_{\rm
p}\lesssim 1h^{-1}$~Mpc, especially for fainter samples (see Figure
\ref{f:vcomp}).

The halo occupation distribution (HOD) specifies the distribution of
the number of galaxies within a (distinct) halo of mass $M$, $P(N|M)$.
It has become a popular tool for interpreting galaxy clustering
(\citealt{Jing98b,Seljak00,Scoccimarro01}; \citealt*{Bullock02,
Yan03}; \citealt{Berlind03,Zehavi05,Kravtsov04,Zheng04,Abazajian05,
Tinker05}), which requires that the first two moments of $P(N|M)$ be
specified to calculate two-point clustering.  In the right panel of
Figure \ref{f:sdss_xi} we show the first moment of this distribution,
the average number of galaxies within a (distinct) halo of mass $M$,
$\langle N(M)\rangle$, for the four halo samples which correspond to
the four luminosity threshold SDSS samples in the left panels.  As
expected, the halo samples corresponding to brighter galaxy samples
reside preferentially in more massive distinct halos.  The halo sample
corresponding to the brightest galaxies ($M_r-5{\rm log}h<-21$) rarely
has more than one halo per distinct halo. All three halo samples
display a gradual roll-off in $\langle N(M)\rangle$ at low mass which
is simply due to scatter in the \Vmax-mass relation, as we select
samples using \Vmax, but plot as a function of mass.  See
$\S$\ref{s:hod} for a more detailed discussion of the HOD associated
with this model.

The good agreement between the observed galaxy correlation function
and samples of halos with our $L-${\Vmax} model, over a range of
luminosities and scales, suggests that the luminosity dependence of
galaxy clustering is due primarily to how galaxies form within dark
matter halos.  This implies that galaxy properties vary as a function
of larger scale environment only insofar as the
halos in which the galaxies reside vary.

\begin{figure}
\plotone{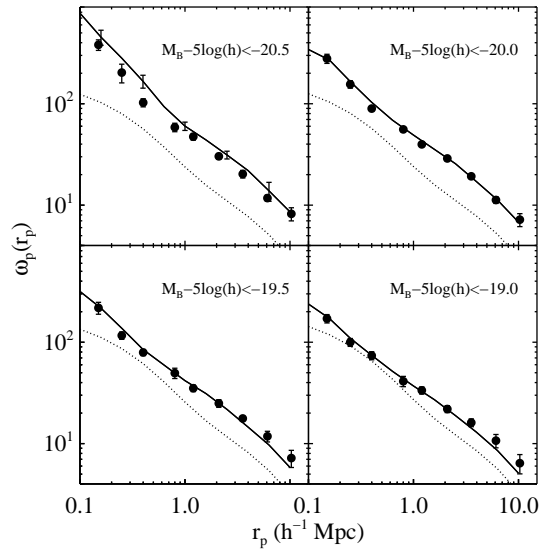}
\vspace{0.5cm}
\caption{Projected two-point correlation function at $z\sim1$ for
DEEP2 galaxies (solid circles) and halos (solid lines), at four
different luminosity thresholds.  We include jack-knife errors,
computed using the eight octants of the simulation cube, on the model
prediction for the brightest sample to demonstrate that they agree
within $1\sigma$.  The excellent agreement on all scales for these
four samples suggests that luminosity-dependent clustering is a result
of two effects: a simple relation between galaxy luminosities and dark
matter halos, and the spatial clustering of the halos.  For
comparison, we include the correlation function of dark matter
particles (dotted lines).}
\label{f:deep2_compare}
\end{figure}

\subsection{Clustering at $z\sim1$}\label{s:deep2}

The DEEP2 Galaxy Redshift Survey \citep{Davis04} has gathered optical
spectra for $\sim50,000$ galaxies at $z\sim1$ using the DEIMOS
spectrograph on the Keck II 10-m telescope.  The survey, recently
completed, spans a comoving volume of $\sim10^6 h^{-3}$ Mpc$^3$,
covering $3$ deg$^2$ over four widely separated fields.  We use the
DEEP2 $B$-band luminosity function of \citet{Willmer05} to compute the
$L-${\Vmax} relation at $z\sim1$.  A Schechter fit to the overall
luminosity function yields $M^\ast_B-5{\rm log}h=-20.73$ and
$\phi^\ast=8.7\times 10^{-3} h^{-3}$ Mpc$^3$ with $\alpha$ fixed at
$\alpha=-1.30$.  A detailed comparison has shown that these values are
consistent with other estimates of the global luminosity function at
$z\sim1$ \citep{Faber05}.

The projected two-point correlation function, $\omega_p(r_p)$, has
been measured for DEEP2 galaxies as a function of luminosity and color
\citep{Coil04a,Coil05a,Coil05b}.  In addition, \citet{Coil05a} has
estimated the two-point cross correlation between galaxies and groups,
and between group centers and group galaxies, based on the group
catalog of \citet{Gerke05}.  For the following comparisons we use the
most recent measurements of $\omega_p(r_p)$ derived from the completed
survey \citep{Coil05b}.

Figure \ref{f:deep2_compare} compares the projected two-point
correlation function for DEEP2 galaxies in four luminosity threshold
samples to the clustering of corresponding dark matter halos.  For the
$M_B-5{\rm log}h<-19.0$, $-19.5$, and $-20.0$ samples we use the L80
simulation, while for the $M_B-5{\rm log}h<-20.5$ sample we use the
L120 simulation to improve statistics.  Slight discrepancies on small
scales for the $M_B-5{\rm log}h<-20.5$ sample may be attributed to
cosmic variance and poisson noise in a sample of this number density,
and in fact our smaller L80 box provides a slighly better match to the
data.  Overall the agreement is excellent on all scales for all four
samples\footnote{Booyah!}.

We would like to stress again that this remarkable agreement is
achieved using the halo distribution in {\it dissipationless}
simulations using a simple, non-parametric relation between galaxy
luminosity and halo circular velocity.  The luminosity-dependent bias
at $z\sim1$ hence seems to be driven entirely by the fact that
brighter galaxies reside in more massive halos, with the
correspondence between halo and luminosity determined by matching the
observed luminosity function to the dark matter halo velocity function.

\subsection{Clustering at $z>2$}

Very little was known about the clustering properties of galaxies at
$z\geq2$ until the advent of the Lyman-break technique
\citep{Steidel93,Steidel96,Steidel99,Steidel03}.  This technique
allows the identification of high-redshift Lyman-break galaxies (LBGs)
by optical photometry alone, using information about the well-defined
region in color-color space that these objects occupy. While simple
color-color cuts allow one to gather large numbers of LBG candidates
with relative ease, it should be kept in mind that this technique is
not perfect.  Accurate completeness numbers are difficult to estimate,
but it is believed that the Lyman-break technique successfully
identifies $80-90\%$ of real LBGs \citep[specific completeness numbers
depend on limiting apparent magnitude, dataset --- e.g., ground
observations or the HST, and sample definition, among other unknowns,
and is often estimated via Monte Carlo simulations of artificial LBGs;
see][]{Adelberger04, Ouchi04a,Lee05}.

The level of contamination, or the fraction of false positives, is
more important when considering the clustering of LBGs because such
objects can artificially dilute or enhance the observed signal.  At
$z\sim3$ the main source of contamination is Galactic stars (4\%) and
high-redshift active galactic nuclei (3\%), as determined by extensive
spectroscopic follow-up \citep{Steidel03}.  At higher redshifts the
situation is less certain, as there has to date been much less
spectroscopic follow-up.  \citet{Ouchi05} estimate the contamination
for their $z\sim4$ sample based on Subaru data at $\sim5$\% from
spectroscopic follow-up of 63 LBG candidates.  LBG candidates
extracted from the GOODS survey avoid Galactic star contamination
thanks to high angular resolution of the HST \citep{Lee05}, although
other sources of contamination remain unquantified.

\begin{figure}
\plotone{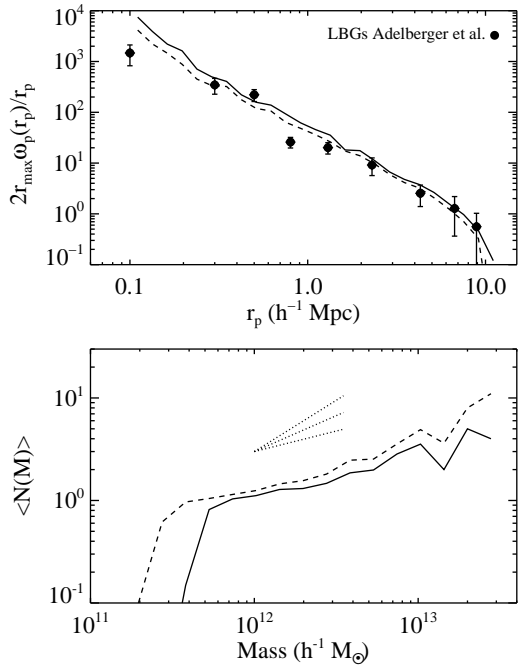}
\vspace{0.5cm}
\caption{Comparison between $z\sim3$ LBGs and dark matter halos in
simulations.  {\it Top panel}: the projected correlation function
measured from $700$ spectroscopically confirmed LBGs
\citep[solid circles,][]{Adelberger03}, compared to halos at the same
number density, $n=4\times10^{-3}h^3$ Mpc$^{-3}$ (solid line) and
halos twice as numerous, $n=8\times10^{-3}h^3$ Mpc$^{-3}$ (dashed
line).  {\it Bottom panel}: The number of halos per distinct halo for
the two halo samples.  The fan (dotted lines) corresponds to slopes of
0.4, 0.7, and 1.0.}
\label{f:z3_xi}
\end{figure}

\begin{figure*}
\plottwo{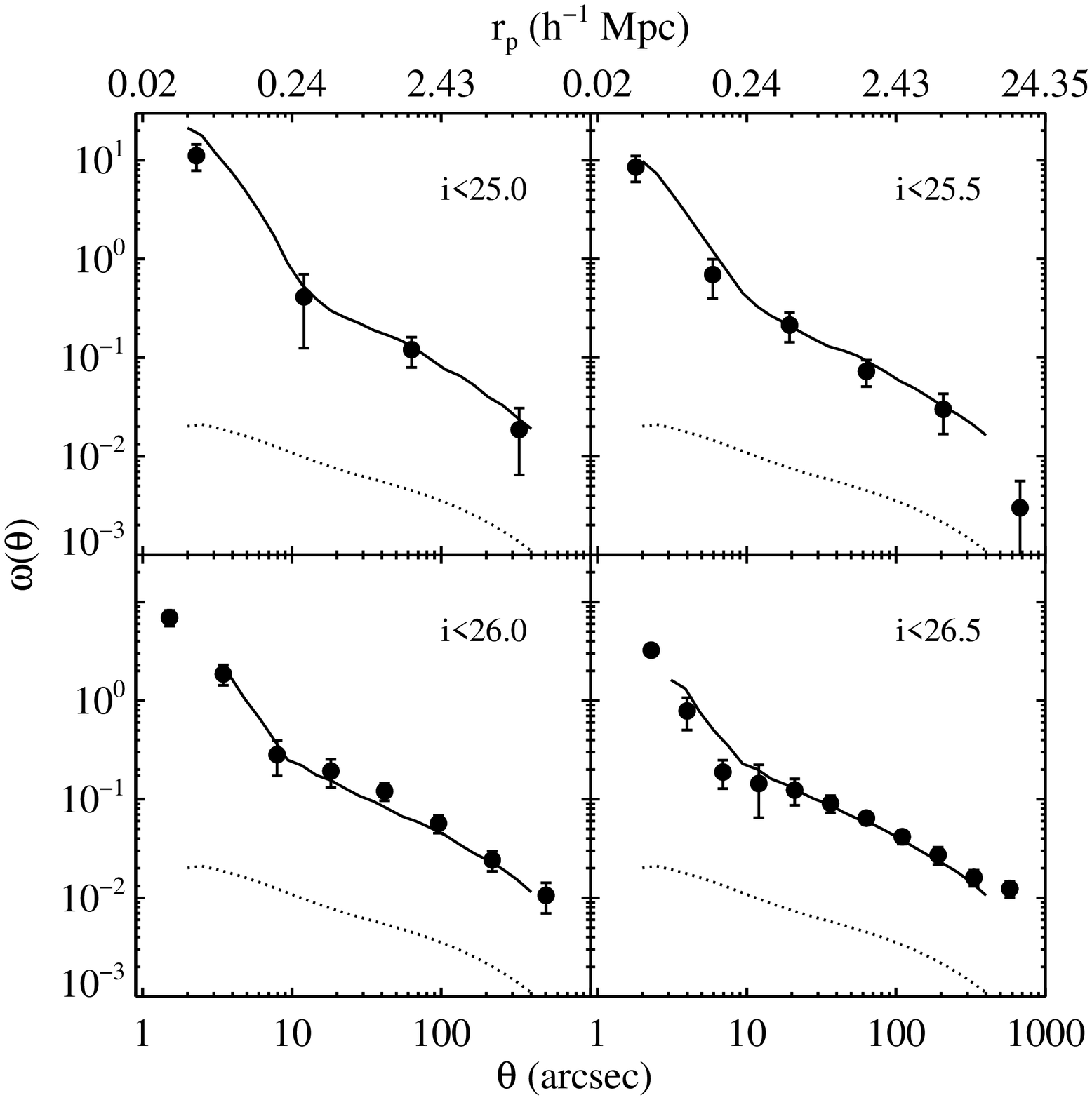}{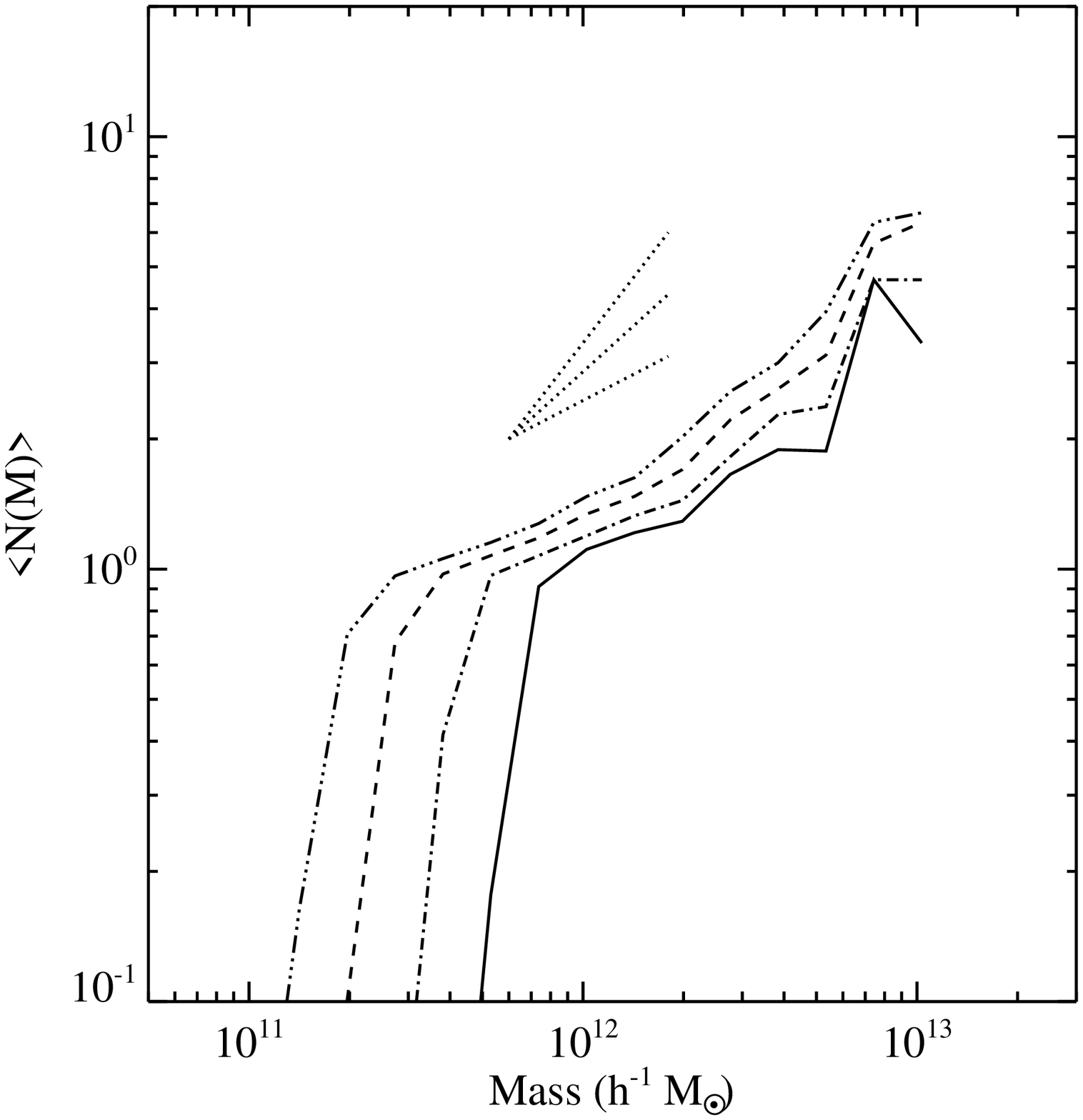}
\vspace{0.5cm}
\caption{{\it Left:} Two-point correlation function of $z\sim4$ LBGs
  derived from Subaru data \citep[solid circles,][]{Ouchi05} compared to
  the correlation function for halos at the same number density (solid
  lines).  The four panels correspond to four different apparent
  magnitude limits, and hence different number densities (see
  Table~\ref{t:summary} for a summary of the Subaru data).  The dotted
  line is $\omega(\theta)$ for dark matter particles at $z\sim4$.
  {\it Right:} Average number of member galaxies for halos
  corresponding to the four Subaru samples at $z\sim4$.  The lines
  correspond to the Subaru samples, from left to right, $i<26.5$,
  $i<26.0$, $i<25.5$, and $i<25.0$.  The fan (dotted lines)
  corresponds to slopes of 0.4, 0.7, and 1.0.}
\label{f:subaru_xi}
\end{figure*}

In this section all quoted number densities have been completeness and
contamination corrected.  If the completeness is well estimated, then
using the corrected number density should mitigate any incompleteness
effects.  However, even if the number density is contamination
corrected, in order to fairly compare the clustering of LBGs to dark
matter halos we must include the clustering properties of these
contaminants.  For simplicity, when estimating the effects of
contamination we will assume the contaminants to have a random
distribution on the sky.  (We need not include additional clustering
effects in the completeness case because in this case the missing
objects are LBGs, which are assumed to have the same clustering
properties as the identified LBGs, but note that this assumption would
break down if the LBG completeness was a strong function of
luminosity.)

Since the early result \citep{Steidel98} that Lyman-Break Galaxies are
a strongly clustered population, there have been numerous attempts to
use their clustering properties to connect these galaxies to their
host dark matter halos \citep[e.g.,][]{Wechsler98,
Jing_Suto98,Adelberger98,Katz99,
Mo99,Wechsler01,Giavalisco01,Bullock02,Scan03}.  There have been two
popular explanations for the properties of these galaxies: some have
speculated that they are a quiescent star-forming population of the
most massive galaxies \citep[e.g.][]{Coles98, Mo99, Giavalisco01},
while others have suggested that they are more related to temporal
events such as merger-driven starbursts
\citep{Lowenthal97,Kolatt99,Somerville01,Scan03}.  The extent to which
either of these scenarios can be ruled out by the clustering data has
been a matter of some debate, depending on the detailed assumptions
that were made in each case (see \citealt{Wechsler01} for a review of
these issues).

\begin{figure}
\plotone{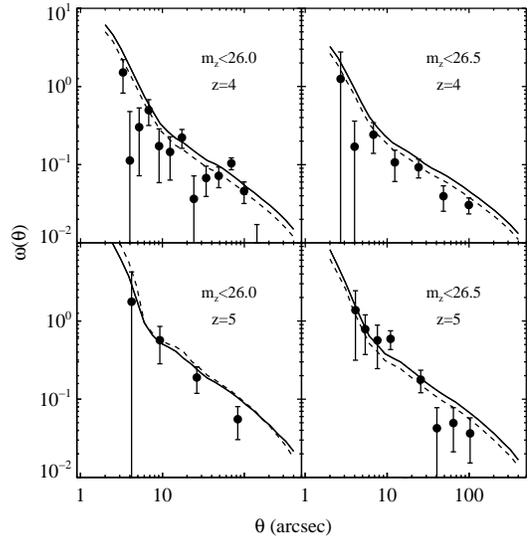}
\vspace{0.5cm}
\caption{Two-point correlation function of LBGs measured using GOODS
data \citep[solid circles,][]{Lee05}, compared to the correlation
function of halos at the same number density (solid lines).  We also
show the effect of a $10\%$ contamination in the identification of
LBGs (dashed lines).  {\it Top panels}: $B$-band dropouts which
correspond to $z\sim4$, at apparent magnitude limits of $m_z<26.0$
(left) and $m_z<26.5$ (right).  {\it Bottom panels}: $V$-band dropouts
corresponding to $z\sim5$ at apparent magnitude limits, $m_z<26.0$
(left), and $m_z<26.5$ (right).  See Table~\ref{t:summary} for a
summary of the GOODS data used here.}
\label{f:goods_xi}
\end{figure}

Our results for LBGs at $z\sim3$ are shown in Figure \ref{f:z3_xi}.
In the top panel we plot the two-point projected correlation function
for $700$ spectroscopically confirmed LBGs \citep{Adelberger03} with
an estimated number density of $n=4\times10^{-3}h^3$ Mpc$^{-3}$ at
$\bar{z}=2.9$.  Note that this sample does not suffer contamination
problems because these LBGs are spectroscopically confirmed.  We plot
$\omega_p$ for halos in units identical to those in \citet[][where
$r_{\rm{max}}$ is the line-of-sight distance through which we count
projected pairs]{Adelberger03}.  We measure $\omega_p(r_p)$ for halos
at two different number densities to illustrate that a factor of two
uncertainty in the LBG number density \citep[a larger uncertainty than
is quoted in][]{Adelberger03} does not alter these results.  The
figure shows that the agreement between the clustering of halos and
LBGs is quite good.

In the bottom panel of Figure \ref{f:z3_xi} we show the halo
occupation of galaxies for the halo samples we associate with $z\sim3$
LBGs.  As expected, the halo sample which has a higher number density
(dashed lines), and a correspondingly lower \Vmax threshold, has a
lower cutoff in $\langle N(M) \rangle$.  The measured halo occupation
implies that most distinct halos are host to a single LBG.  $\langle
N(M) \rangle$ rises above $2$ only for the most massive halos
($\gtrsim10^{13} h^{-1} M_{\Sun}$), which are very rare at $z\sim3$.

At still higher redshifts the clustering of LBGs has recently been
measured in two separate datasets: by \citet{Ouchi05} using Subaru
data and by \citet{Lee05} using the HST GOODS survey.  These authors
detect a strong departure from a power law in the angular two-point
correlation function of LBGs.  When parameterized in the framework of
the halo model, this excess small scale ($\theta<10''$)  power is
attributed to multiple galaxies within a distinct halo, i.e. the
`1-halo' term.  We now compare our $L-${\Vmax} model to these data.

The left panel of Figure \ref{f:subaru_xi} compares the angular
correlation function of LBGs from \citet{Ouchi05} to the angular
correlation function of halos for four different apparent magnitude
thresholds, at $z\sim4$.  If the identified LBGs occupy a relatively
narrow range of redshifts, as expected, then these apparent magnitude
cuts should correspond to absolute magnitude limits. The agreement is
again remarkably good on all the scales we probe. Our non-parametric
$L-${\Vmax} model captures the luminosity dependence of LBG
clustering, on both large and small scales, correctly predicts the
scale  of the small-scale upturn in $\omega(\theta)$, and the fact
that this scale decreases for increasing number density.  The latter
trend is due to a decrease in the typical mass (and size) of the
distinct halos hosting LBGs as the number density is increased.  We do
not plot the correlation function for halos below $r_p\approx 50
h^{-1}$ kpc since the halo finder we use does not find halos closer
than this separation.

Connecting with previous work, we show in the right panel of Figure
\ref{f:subaru_xi} the halo occupation, $\langle N(M) \rangle$, for the
halo samples which correspond to the $z\sim4$ Subaru samples.
Increasingly brighter LBG samples reside in preferentially larger mass
distinct halos.  Furthermore, since $\langle N(M) \rangle<2$ for
$M<10^{13} h^{-1} M_{\Sun}$, our $L-${\Vmax} model implies that the
majority of these highest-redshift LBGs live alone in distinct dark
matter halos.  This picture is in qualitative agreement with previous
analysis of the LBG clustering in the framework of halo model
\citep{Lee05,Ouchi05}.

Figure \ref{f:goods_xi} compares the clustering of LBGs from GOODS
data at $z\sim 4$ (top panels) and $z\sim5$ (bottom panels) to the
clustering of halos at similar redshifts and number densities.  Here
again the agreement is very good,  even at $z\sim 5$. The model
somewhat overpredicts the clustering of the fainter sample at $z=4$.
In this plot we also show the effect of a contaminated LBG sample,
i.e. we assume that $10\%$ of identified LBGs are actually interlopers
with a random distribution on the sky.  The effect of contamination
scales $\omega(\theta)$ by $(1-c)^2$, where $c$ is the contamination
fraction.  A $10\%$ contamination reduces the angular correlation
function by $\approx 20\%$ and results in better agreement for the
fainter GOODS samples (right panels) which plausibly have higher
contamination than the brighter samples.  We note that $10\%$
contamination is likely an upper limit and was included here for
illustration purposes.  The detailed spectroscopic follow-up of these
high redshift samples have yet to yield direct accurate estimates of
contamination fractions.

The straightforward implication of the presented comparisons is that
the data is consistent with, and one could argue supports, the picture
in which most LBGs are the central galaxies in their host halos with
luminosity tightly related to the halo circular velocity and mass.
Most LBGs have no neighbors within the same halo. However,  a fraction
of them do and it is this fraction that is responsible for the strong
upturn in the correlation function at small scales. By accurately
reproducing both the small-scale upturn in $\omega(\theta)$ and the
large-scale clustering, our model accurately predicts not only the
correct distinct halos to associate with LBGs (the `2 halo term' in
halo model jargon) but also the number of LBGs within a distinct halo
(the corresponding `1 halo term').

\begin{deluxetable}{llllll}
\tablecaption{Summary of Samples} \tablehead{\colhead{Data} &
\colhead{Defn.\tablenotemark{a}} & \colhead{$\bar{z}$
\tablenotemark{b}} & \colhead{$n/10^{-3}$} & \colhead{\Vmax
\tablenotemark{c}} \\ \colhead{} & \colhead{} & \colhead{} &
\colhead{$h^{3}$ Mpc$^{-3}$} & \colhead{km s$^{-1}$}} \startdata  SDSS
& $M_r<-18$ & $0.04$ & $27.0$ & $110$ \\ SDSS & $M_r<-19$ & $0.06$ &
$15.0$ & $130$ \\  SDSS & $M_r<-20$ & $0.06$ & $6.0$ & $180$ \\  SDSS
& $M_r<-21$ & $0.15$ & $1.1$ & $310$ \\  DEEP2 & $M_B<-19.0$ & $0.87$
& $13.0$ & $150$ \\ DEEP2 & $M_B<-19.5$ & $0.92$ & $8.4$ & $170$ \\
DEEP2 & $M_B<-20.0$ & $0.98$ & $4.9$ & $200$ \\ DEEP2 & $M_B<-20.5$ &
$0.99$ & $2.5$ & $250$ \\ Adelberger & $U_{n}GR$ colors & $2.9$ &
$4.0$ & $207$ \\ Subaru & $i<25.0$ & $4$ & $0.8$ & $265$ \\ Subaru &
$i<25.5$ & $4$ & $1.9$ & $225$ \\ Subaru & $i<26.0$ & $4$ & $3.8$ &
$191$ \\ Subaru & $i<26.5$ & $4$ & $6.4$ & $174$ \\ GOODS &
$m_z<26.0$, $B$-drop & $4$ & $2.7$ & $205$ \\ GOODS & $m_z<26.5$,
$B$-drop & $4$ & $4.5$ & $185$ \\ GOODS & $m_z<26.0$, $V$-drop & $5$ &
$1.5$ & $200$ \\ GOODS & $m_z<26.5$, $V$-drop & $5$ & $2.6$ & $180$ \\
\enddata \tablenotetext{a}{Absolute SDSS and DEEP2 magnitudes are in
units of $M-5\rm{log}(h)$.}  \tablenotetext{b}{Mean redshift of
sample; for Subaru and GOODS data, $\bar{z}$ is based on Monte Carlo
simulations of artificial LBGs.}  \tablenotetext{c}{\Vmax such that
$n(>V_{\rm{max}})=n_{\rm{sample}}$ for halos at the simulation output
closest to $\bar{z}$. }
\label{t:summary}
\end{deluxetable}

\section{Dependence of Halo Occupation on number density 
and redshift }\label{s:hod}

In this section we explore the redshift and number density dependence
of the halo occupation distribution (HOD), the key ingredient of the
halo model, in our \Vmax$-L$ assignment scheme. We compare our results
to previous studies which fit the halo occupation to observational
clustering data. The HOD, or the probability distribution for a
distinct halo of mass $M$ to host $N$ galaxies, provides a relatively
clean and simple framework for interpreting clustering data. The
two-point correlation function in the halo model depends on the first
and second moments of the HOD ($\langle N\rangle$ and $\langle
N(N-1)\rangle$, respectively).

The first moment of the HOD can be separated into two components
\citep[e.g.,][]{Kravtsov04}:

\begin{equation}\label{eqn:nall}
\langle N(M) \rangle = \langle N(M) \rangle_{\rm{cen}} + \langle N(M)
\rangle_{\rm{sat}},
\end{equation}
where $\langle N(M) \rangle_{\rm{cen}}$ and $\langle N(M)
\rangle_{\rm{sat}}$ are the number of central and satellite galaxies
(i.e., subhalos), respectively.  The central galaxy term is a step
function rising to $\langle N(M) \rangle_{\rm{cen}}=1$ above a minimum
distinct halo mass, $M_{\rm{min}}$.  \citet{Kravtsov04} show that
distribution of the subhalo occupation at fixed $M$ is well-modeled by
a Poisson distribution; this continues to hold well in the simulations
presented here.  Although some authors model the cutoff at
$M_{\rm{min}}$ to take into account scatter between the observable and
host halo mass, we model this term as a step-function for simplicity,
and note that this simplification will only impact the $M_{\rm{min}}$
parameter.

\begin{figure}[t!]
\plotone{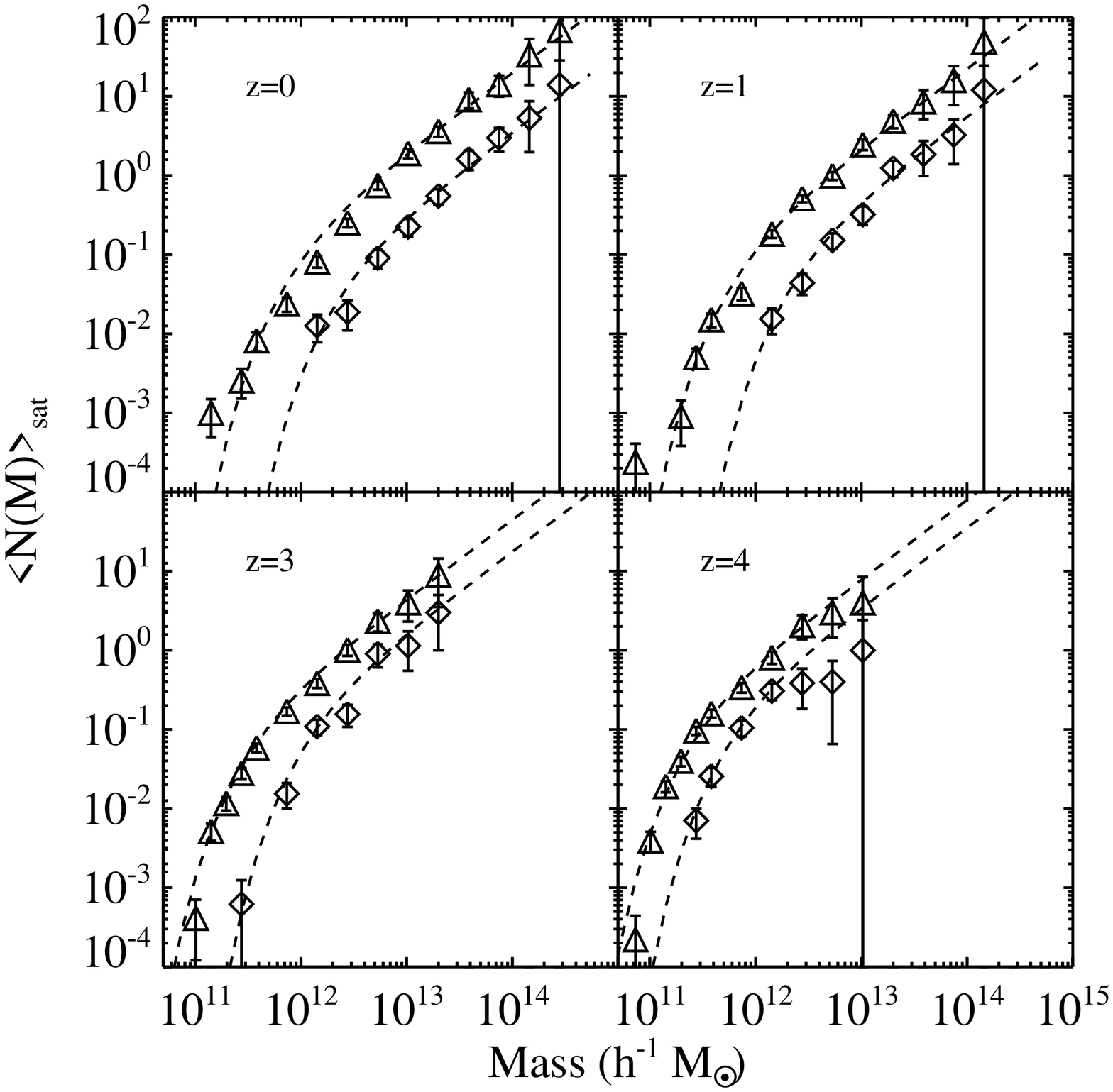}
\vspace{0.5cm}
\caption{$\langle N(M)\rangle_{\rm{sat}}$ as a function of number
density and redshift. Diamonds designate halo samples at
$n=4.0\times10^{-3}$ $h^3$ Mpc$^{-3}$, triangles are for samples with
$n=20.0\times10^{-3}$ $h^3$ Mpc$^{-3}$.  The dashed lines corresponds
to Eqn.~\ref{eqn:nsat_exp}, with the best-fit parameters listed in
Table~\ref{t:hod_ev}.}
\label{f:hod_nsat_ev}
\end{figure}

The satellite term, $\langle N(M)\rangle_{\rm{sat}}$, can be described
by a power-law function, $\langle N(M) \rangle_{\rm{sat}} \propto
M^{\alpha}$, for large distinct halo masses.  \citet{Kravtsov04} and
\citet{Zheng05} find that $\alpha \approx 1$ for subhalos and galaxies
identified in cosmological simulations within massive distinct halos,
over a wide range of number density thresholds (corresponding to
luminosity threshold) cuts. For small distinct halo masses,  $\langle
N(M) \rangle_{\rm{sat}}$ ``rolls-off'' faster than the power law.
\citet{Kravtsov04} describe the ``roll-off'' by:
\begin{equation}\label{eqn:nsat_krav}
\langle N(M) \rangle_{\rm{sat}} = (M/M_1-C)^\alpha .
\end{equation}
\citet{Tinker05} propose an alternative exponential form for the
``roll-off'':
\begin{equation}\label{eqn:nsat_exp}
\langle N(M) \rangle_{\rm{sat}} =
\frac{M}{M_1}\exp\left({-\frac{M_{\rm{cut}}}{M}}\right),
\end{equation}
where $M_{\rm{cut}}$ and $M_1$ are free parameters, this time with the
asymptotic slope fixed at $\alpha=1$.  In fact, we find that
$M_{\rm{cut}}$  and $M_1$ are tightly correlated such that
\begin{equation}\label{eqn:mcut}
  \log(M_{\rm{cut}}) = 0.76 \log(M_1) + 2.3,
\end{equation}
for the full range of number densities and redshifts explored here,
thereby reducing Equation~\ref{eqn:nsat_exp} to a one-parameter
function.  As shown in  Figure \ref{f:hod_nsat_ev}, this form is an
excellent fit to the number of satellite halos in our simulations over
a large range of redshifts and number densities.  We have explicitly
checked the asymptotic slopes of $\langle N(M) \rangle_{\rm sat}$
where possible and find that they are consistent with $\alpha=1$, in
agreement with previous theoretical predictions \citep{Kravtsov04}.

The exponential functional form of Equation~\ref{eqn:nsat_exp} is
preferable to Equation~\ref{eqn:nsat_krav} because for a given choice
of parameters $M_1$ and $C$, $\langle N(M) \rangle_{\rm{sat}}$ can
become negative using the latter.  For example, best-fit values for
these parameters for the $z\sim0$, $n=20.0\times10^{-3}$ $h^3$
Mpc$^{-3}$ sample (triangles in the top left panel of Figure
\ref{f:hod_nsat_ev}) implies that $\langle N(M) \rangle_{\rm{sat}}$
will be negative for $M\sim2\times10^{12}$ $h^{-1} M_{\Sun}$, which is
not in agreement with the results of simulations. For this reason we
choose to use Equation ~\ref{eqn:nsat_exp} to describe the functional
form of $\langle N(M) \rangle_{\rm{sat}}$.

Note that the presence of the ``roll-off'' is important, and
neglecting it can bias the best-fit HOD parameters.  \citet{Zehavi05}
use
\begin{equation}\label{eqn:nsat_pow}
\langle N(M) \rangle_{\rm{sat}} = (M/M_1)^\alpha
\end{equation}
to model the clustering of SDSS galaxies at $z\sim0$ and find that
$\alpha$ increases for galaxies of increasing luminosity.  However, in
the presence of the roll-off at small $M$, brighter samples will have
on average fewer satellites and hence $\langle N(M)
\rangle_{\rm{sat}}$ will be dominated by the steeper exponential
cut-off.  A power-law fit is hence expected to yield artificially
steep slopes.  Indeed, when fitting a pure power law without this
roll-off to $\langle N(M) \rangle_{\rm{sat}}$ in the simulations, we
find slopes closer to $\alpha\approx1.5$.

\begin{figure}[t!]
\plotone{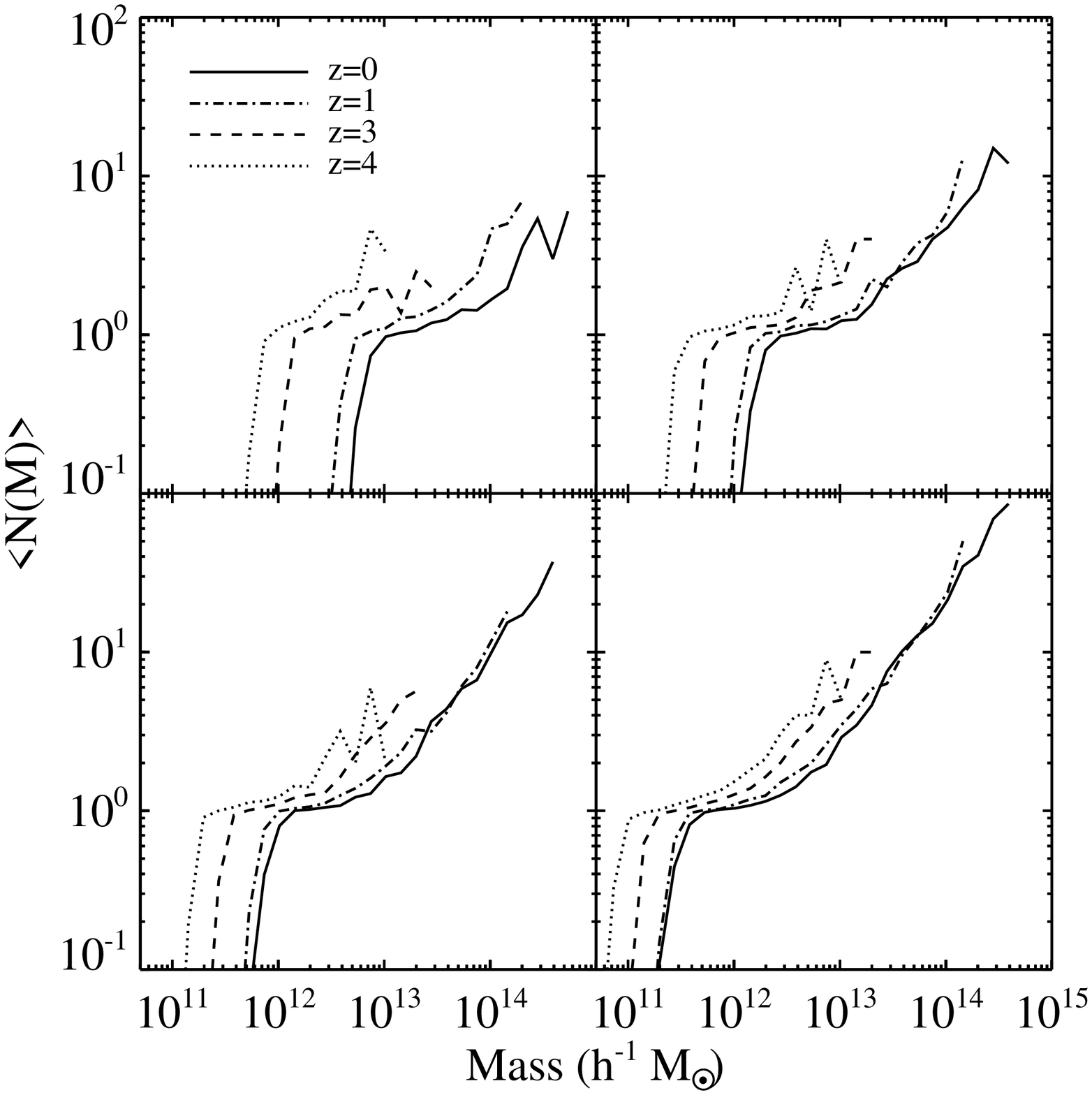}
\vspace{0.5cm}
\caption{Evolution of the halo occupation, $\langle N(M) \rangle$
from $z\sim4$ to $0$ at fixed number density (clockwise from top left:
$n=1.0$, $4.0$, $20.0$, and $8.0$, in units of $10^{-3} h^3$
Mpc$^{-3}$).}
\label{f:hod_ev}
\end{figure}

In Table~\ref{t:hod_ev} we list the best-fit HOD parameters for halos
in our simulations over a wide range of number densities and
redshifts.  The corresponding halo occupations are plotted in Figure
\ref{f:hod_ev}, while Figure \ref{f:hod_nsat_ev} shows the
contribution from subhalos (satellites) alone.

\begin{deluxetable}{ccccccc}
\tablecaption{HOD parameters and derived quantities}
\tablehead{\colhead{n/$10^{-3}$} & \colhead{$z$} &
\colhead{$\log(M_{\rm{min}})$} & \colhead{$\log(M_{1})$} &
\colhead{$\log(M_{\rm{cut}})$} & \colhead{$\langle N \rangle$} &
\colhead{$\log(\langle M \rangle)$} \\ \colhead{$h^3$ Mpc$^{-3}$} &
\colhead{} & \colhead{} & \colhead{} & \colhead{} & \colhead{} &
\colhead{} } \startdata $1.0$ & $0$ & $12.8$ & $14.0$ & $13.1$ &
$1.15$ & $13.6$ \\ $1.0$ & $1$ & $12.7$ & $13.7$ & $12.7$ & $1.16$ &
$13.2$ \\ $1.0$ & $3$ & $12.1$ & $12.0$ & $12.0$ & $1.15$ & $12.5$ \\
$1.0$ & $4$ & $11.8$ & $12.4$ & $12.1$ & $1.16$ & $12.2$ \\ \\
\hline\\ $4.0$ & $0$ & $12.3$ & $13.5$ & $12.4$ & $1.22$ & $13.4$ \\
$4.0$ & $1$ & $12.1$ & $13.3$ & $12.4$ & $1.14$ & $12.9$ \\ $4.0$ &
$3$ & $11.7$ & $12.8$ & $12.1$ & $1.08$ & $12.2$ \\ $4.0$ & $4$ &
$11.5$ & $12.5$ & $11.8$ & $1.08$ & $11.8$ \\ \\ \hline \\ $8.0$ & $0$
& $12.0$ & $13.1$ & $12.4$ & $1.27$ & $13.4$ \\ $8.0$ & $1$ & $11.9$ &
$13.0$ & $12.3$ & $1.17$ & $12.8$ \\ $8.0$ & $3$ & $11.5$ & $12.6$ &
$11.9$ & $1.08$ & $12.0$ \\ $8.0$ & $4$ & $11.2$ & $12.3$ & $11.6$ &
$1.08$ & $11.7$ \\ \\ \hline \\$20.0$ & $0$& $11.5$ & $12.7$ & $12.0$
& $1.32$ & $13.4$ \\ $20.0$ & $1$& $11.5$ & $12.7$ & $11.9$ & $1.20$ &
$12.7$ \\ $20.0$ & $3$& $11.1$ & $12.4$ & $11.6$ & $1.09$ & $11.8$ \\
$20.0$ & $4$& $10.9$ & $12.1$ & $11.5$ & $1.07$ & $11.5$ \\ \enddata
\tablecomments{All quoted masses are in units of $h^{-1} M_\Sun$.}
\label{t:hod_ev}
\end{deluxetable}

With the HOD in hand we can derive several useful quantities.  The
average number of halos per distinct halo,
\begin{equation}
\langle N \rangle = \frac{\int^\infty_{M_{\rm{min}}} \, \langle
N(M)\rangle n_{\rm{halo}}(M) dM} {\int^\infty_{M_{\rm{min}}}
n_{\rm{halo}}(M) dM}
\end{equation}
and the average mass of distinct halos (member number weighted),
\begin{equation}
\langle M \rangle = \frac{\int^\infty_{M_{\rm{min}}} \, \langle
N(M)\rangle n_{\rm{halo}}(M) M dM} {\int^\infty_{M_{\rm{min}}}
n_{\rm{halo}}(M) dM}
\end{equation}
are useful when considering a `typical' halo in a given sample.  These
two quantities are included in Table~\ref{t:hod_ev}.  Note that when
computing these quantities we use the halo occupation measured
directly from simulations (shown in Figure \ref{f:hod_ev}) rather than
the best-fit halo occupation implied by equations ~\ref{eqn:nall} and
~\ref{eqn:nsat_exp}.

The redshift evolution of $\langle N(M) \rangle$ for samples of fixed
number density reveals several interesting trends. The minimum mass of
the sample, or the location of the step in the mean occupation number,
is decreasing with increasing redshift, reflecting evolution of the
halo mass function. At higher redshifts, the halos hosting multiple
satellites become more rare, so that in most samples the high-$M$
power-law tail of $\langle N(M) \rangle_{\rm sat}$ is not
present. This trend is apparent in Figure \ref{f:hod_nsat_ev} which
shows the redshift evolution of $\langle N(M) \rangle_{\rm sat}$ for
halo samples of different number densities.  At the same time, the
``shoulder'' (or the region between the step and the power-law tail)
in $\langle N(M) \rangle$ becomes shorter and not as flat at higher
$z$ (Fig.~\ref{f:hod_ev}), reflecting the increasing fraction of
relatively small-mass halos hosting more than a single central
galaxy. We choose to characterize the extent of the shoulder via the
ratio $M_1/M_{\rm{min}}$.  As can be seen by consulting
Table~\ref{t:hod_ev}, this ratio (and hence the shoulder) decreases
both with decreasing number density and increasing redshift.  As we
will discuss more thoroughly below, we believe that it is the extent
of the shoulder which is primarily responsible for the upturn in the
correlation function on small scales.  Hence samples with a lower
number density and/or at higher redshift (and hence a smaller
shoulder) should have a more significant upturn in $\omega_p(r_p)$ on
small scales, as is observed.

The HOD parameters we derive are in good qualitative agreement with
observations.  At $z\sim0$, \citet{Zehavi05} fit a different
functional form for $\langle N(M) \rangle_{\rm{sat}}$ (namely
Eqn.~\ref{eqn:nsat_pow}), and hence a direct comparison is
complicated.  However, $M_{\rm{min}}$ is unaffected by the functional
form for $\langle N(M) \rangle_{\rm{sat}}$, and here the values quoted
in Table~\ref{t:hod_ev} are in excellent agreement with
\citet{Zehavi05}.  The trend of $M_1$ to decrease with increasing
number density is also in agreement with the derived results from
observations.

A direct comparison to the HOD derived in the literature for LBGs is
again complicated. The parameters given in Table~\ref{t:hod_ev} are
{\it predicted} by our simulations, while most analyses {\it assume} a
pure power-law form for the overall halo occupation and fit for the
parameters in the framework of the halo model
\citep[][]{Hamana05,Lee05,Ouchi05}. The assumed power-law form does
not describe HOD in our simulations well. This is an example
illustrating that HOD fitting is in general not unique --- different
functional forms can simultaneously match the number density and
2-point clustering of a sample.  Because derived quantities such as
$\langle N \rangle$ and $\langle M \rangle$ are entirely dependent on
the form of the HOD, it is critical to realize that these derived
quantities are also not unique.  Despite the different forms for
$\langle N(M) \rangle$, we find here similar qualitative trends for
the average distinct halo mass of LBGs, namely that fainter samples
(corresponding to higher number densities) have smaller $\langle M
\rangle$, as is seen in \citet{Ouchi05} and \citet{Lee05}.  As emphasized by
\citet*{Bullock02}, this degeneracy may be partially broken by looking
at the smallest scale clustering information available (typically
parameterized as the close-pair counts); a comparison with the most
recent data of this sort from $z\sim 0-1$ is being pursued for a
similar model to ours by Berrier et al (in preparation).

As was shown in $\S$\ref{s:appr}, selecting subhalos according to \Vin
results in an enhanced clustering signal compared to using \Vnow,
especially on small scales and at low redshifts, because there are
more subhalos in the \Vin selected sample.  To illustrate this
directly, we fit Eqn.~\ref{eqn:nsat_exp} to satellites selecting in
these two different ways.  We find that the effect of using \Vin over
\Vnow, insofar as there are more satellites at fixed distinct halo
mass, is small for low masses, and gradually increases toward higher
masses. In other words, the difference in $\langle N(M)\rangle$ is
small near the exponential cutoff while {\Vin}-selected samples show
steeper $\langle N(M)\rangle$ with increasing $M$ for higher masses.

To understand this trend consider the following. In order for there to
exist a sizable difference between  \Vin and \Vnow, a subhalo must
orbit within a distinct halo for a sufficient amount of time such that
tidal stripping can reduce \Vmax.  Dynamical friction, which depends
on the relative masses of subhalos to distinct halos, will be much
stronger in lower distinct halo masses because there exists a minimum
halo mass corresponding to our \Vmax threshold.  In other words, as
one considers smaller distinct halo masses for a given \Vmax
threshold, the typical subhalo mass cannot be arbitrarily close to the
host mass because such halos would suffer efficient dynamical friction
and would merge quickly without significant evolution of {\Vmax}.  At
higher distinct halo masses, dynamical friction is not efficient for
the typical subhalo since now the typical subhalo is much less massive
than the distinct halo, and hence a subhalo can exist within a
distinct halo for a sufficient amount of time for tidal stripping to
have a significant effect on \Vmax.

\section{Discussion}\label{s:disc}

Following the reliable identification of subhalos in $N$-body
simulations, there has been a concerted effort to identify such
subhalos with observed galaxies.  A persistent problem with this
approach is that subhalos selected using the present-day mass or
circular velocity tend to have radial distributions within their
larger distinct halos that are shallower than those exhibited by
observed galaxies \citep{Gao04a,Diemand04,Nagai05,Maccio05}.  However,
\citet{Nagai05} find that the observed radial profile of galaxies can
be reproduced by selecting subhalos using stellar mass, rather than
the present day total mass.  We argue that the best proxy for the
stellar mass in dissipationless simulations is the current maximum
circular velocity for distinct halos and the circular velocity at the
time of accretion, \Vin, for subhalos.

We have shown that this simple, non-parametric model which relates
galaxy luminosities to the maximum circular velocity of dark matter
halos, \Vmax, by preserving the observed galaxy luminosity function,
accurately reproduces the observed luminosity-dependent clustering of
galaxies over projected separations $0.1<r_p/(h^{-1}$ Mpc$)<10.0$,
from $z\sim 5$ to $z\sim 0$. The only assumption of the model is that
there is a monotonic relation between galaxy luminosity and halo
circular velocity. To assign luminosities we use the maximum circular
velocity at the time of accretion, \Vin, for subhalos, and the maximum
circular velocity at the time of observation, \Vnow, for distinct
halos.  This ingredient is crucial for accurately reproducing the
observed clustering of galaxies at $z\sim 0$.

The success of the model has several important implications.  First,
our results indicate that dissipationless simulations do not suffer
from significant overmerging because they reproduce both the amplitude
and shape of the correlation function and the underlying HOD quite
accurately. If our dissipationless simulations were missing a
significant fraction ($\gtrsim 40\%$) of objects in groups and
clusters, as suggested by some recent studies
\citep{Gao04a,Diemand04}, both the amplitude and shape of the
correlation function in these simulations would be grossly
incorrect.  We also note that recent analysis of the observed
galaxy-mass cross correlation function disfavors the existence of a
large population of satellites without associated dark matter halos
\citep{Mandelbaum05}.

In order to assess the quantitative impact of a fraction of ``orphan''
galaxies --- satellite galaxies with no identifiable subhalo
\citep{Gao04a} --- in our model, we perform the following test.  We
artificially increase the fraction of subhalos in such a way that the
subhalo occupation function, $\langle N(M)\rangle_{sat}$ increases in
amplitude while maintaining the same shape.  In order to match the
observed galaxy luminosity function, we simulatenously increase the
overall \Vmax threshold for a given sample, such that the number
density does not change.  We find that an orphan fraction $>10$\% is
inconsistent with the correlation functions of the SDSS galaxy samples
at $z\sim0$ for all luminosities (for this cosmological model and the
assumption of no scatter between $L$ and \Vin).

Our results also imply that the central assumption of the luminosity
assignment model --- the tight, monotonic relation between galaxy
luminosity and halo circular velocity --- likely exists for the
observed galaxies.  Such a relation is indeed expected
\citep[e.g.,][]{Mo98} for isolated galaxies, but our results indicate
that this is true globally for galaxies of different types and for a
wide range of redshifts.  We argue that for subhalos the dissipation
should result in a centrally condensed, tightly-bound stellar system
which would stabilize the halo circular velocity against tidal
heating. Stellar mass and luminosity of galaxies should therefore
correlate with the circular velocity of subhalos at the time they are
accreted, before significant tidal evolution takes place. Given that
we match luminosity at a particular epoch to the circular velocity at
different epochs (the epoch of observation for distinct halos, and the
epoch of accretion for subhalos), a subtle implication of our scheme
is that the relation between luminosity and \Vmax does not evolve
strongly with time --- a result which may have been anticipated by the
lack of scatter in the Tully--Fisher relationship in different
environments.

The corollary is then that the clustering of a particular galaxy
sample is largely determined by the clustering of halos and subhalos
that host the galaxies. Clustering of halos and subhalos is governed
by gravitational dynamics \citep[e.g.,][and discussion
therein]{Kravtsov99b,Zentner05}, while the particular subset of halos
that host galaxies in a given sample is determined by the relations
between observable galaxy properties and properties of their host dark
matter halos {\it and} selection criteria used to define the sample.
In the case of the galaxy luminosity, the relation with halo circular
velocity appears to be particularly tight.

The model agreement with the clustering properties of the LBG
population at $3<z<5$ is perhaps more surprising, given that the LBG
selection criteria are significantly more complicated than those of
the luminosity-selected samples at $z\leq 1$.  The success of our
$L-$\Vmax model indicates that the rest-frame LBG luminosity is likely
to be tightly related to the halo circular velocity (and hence total
mass). This would, in turn, suggest that LBGs are fair tracers of the
overall halo population rather than a special subset of halos, such as
halos undergoing mergers or collisions.  Although we have not
explicitly investigated the latter scenarios (i.e., mergers or
collisions), it seems unlikely that the observed LBG clustering would
be well matched by a model in which the effective duty cycle for the
LBGs is small and most LBGs are associated with minor mergers or
small-mass collisions.  We note, however, that differences in the
clustering properties and inferred host masses for the so-called
``massive halo'' model and the ``collisional starburst'' model may
have been exaggerated --- in the former case because the massive
subhalos were ignored, and in the latter because the efficiency of
star formation in mergers was probably overestimated
\citep[e.g.,][]{Cox05}.

A key success of the simulations and our model is the correct
description of the small-scale upturn in the correlation function.  At
$z\sim 0$, the upturn is detected for bright galaxies
\citep{Zehavi04}, while the correlation function of faint objects is
very close to a power law \citep{Zehavi05}. At $z\gtrsim 1$, the
upturn is more pronounced and is now unambiguously detected both in
the DEEP2 data and in LBG samples at $z\sim 3-5$, as predicted from
the halo model arguments \citep{Zheng04} and cosmological simulations
\citep{Kravtsov04}. The trend towards more pronounced deviation from
the power-law form of CF at higher $z$ appears to be due to a couple
different factors.

The accretion rate of subhalos is substantially higher at high $z$, so
that the abundance of subhalos with masses close to the threshold mass
of the sample, $M_{\rm min}$, is also larger on average (see
\S~\ref{s:hod}).  This makes the ``shoulder'' of the HOD of high-$z$
objects shorter and not as flat, increasing the number of
small-separation pairs between central galaxies and satellites and
between satellites themselves in {\it small-mass halos}. This also
makes the width of the HOD  at masses close to $M_{\rm min}$ wider and
closer to the Poisson  distribution. Although the overall fraction of
subhalos in a given sample actually {\it decreases} with increasing
redshift (see Fig.~\ref{f:cumn_z}), the average mass of the distinct
halos hosting most of subhalos is decreasing as well. The contribution
of the pairs of objects within the same halo to the 1-halo term of the
correlation function comes then predominantly from the smallest halos
in the sample and the 1-halo term is considerably ``peakier'' than at
lower redshifts, where halos of a wide range of masses contribute.  We
argue therefore that the more pronounced upturn in the correlation
function at higher redshifts and for rarer objects at a given epoch is
primarily due to the increased fraction of subhalos in host halos with
masses closer to the threshold mass of the sample.

In the present work, we have not included any scatter in matching
luminosity and velocity relations, and it is interesting that making
this approximation works so well.  Some scatter in the $L-${\Vmax}
relation should certainly be expected given the scatter in the
observed Tully-Fisher and Faber-Jackson relations.  This scatter
likely depends on band, galaxy properties, and the extent to which
internal dust extinction has been corrected for.  For our purposes,
the main effect of adding scatter in this relationship to our model
would be at the highest luminosities.  The insensitivity at low
luminosities arises because 1) the luminosity function of galaxies is
rather flat at $L<L_{\ast}$ so that the scatter results in almost
equal numbers of objects scattering in and out of the sample, and 2)
the bias of halos corresponding to these luminosity cuts
($L<L_{\ast}$), is only a weak function of mass and luminosity both in
our simulations and in observations \citep{Tegmark04}, so changes in
the mass distribution of galaxies do not affect clustering
significantly.  In contrast, at higher luminosities, both the
luminosity function and the bias dependence on mass are steep, and
adding scatter serves to decrease the average clustering and makes the
luminosity a steeper function of mass. \citet{Tasitsiomi04} find that
some scatter is required to match the galaxy-mass correlations, but
the conclusion is reached using \Vnow for subhalos, which we expect to
exhibit more scatter with luminosity than \Vin. There is, however, an
interesting tension between constraints provided by the observed
galaxy-galaxy and galaxy-mass correlation functions for the most
massive halos (cf. Fig.~\ref{f:xigm}): the former constrains the
scatter for bright galaxies to be small, while the latter appears to
require significant amount of scatter.  We postpone a full discussion
of these issues including combined constraints to future work.

\vspace{1em}

\section{Summary}\label{s:conc}

Our main results and conclusions are as follows:

\begin{itemize}

\item[1.] We show that a simple, non-parametric model which
monotonically relates galaxy luminosities to the maximum circular
velocity of dark matter halos, \Vmax, by preserving the observed
galaxy luminosity function, accurately reproduces the observed
luminosity-dependent clustering of galaxies over projected separations
$0.1<r_p/(h^{-1}$ Mpc$)<10.0$, through most of the evolution of the
universe from $z\sim 5$ to the present.

\item[2.] The key to the success of the model is our luminosity
assignment scheme, in which we use the maximum circular velocity at
the time of accretion, {\Vin}, for subhalos and the maximum circular
velocity at the time of observation, {\Vnow}, for distinct halos.  We
argue that for subhalos in dissipationless simulations, {\Vin}
reflects the luminosity and stellar mass of the associated galaxies
better than the circular velocity at the epoch of observation, {\Vnow}

\item[3.] Our simulations, coupled with the above luminosity
assignment scheme, correctly reproduce the small-scale deviation of
the correlation function from the power-law form (the upturn) in
observed luminosity-selected samples at different redshifts.  The
deviation is, in general, more pronounced at high redshifts.  We
attribute this trend to the increased fraction of satellite galaxies
in host halos with masses close to the threshold mass of the sample.

\item[4.] Our luminosity assignment model applied to the simulations
at $z\sim 3-5$ reproduces the observed shape and amplitude of the
two-point correlation function of Lyman-break galaxies on both large
and small scales.  This suggests that LBGs are fair tracers of the
overall halo population at these epochs and that their luminosity can
be simply related to the circular velocity (and hence mass)  of their
dark matter halos. Although our model implies that most ($\sim 90$\%)
of the LBGs are central galaxies in distinct halos,  a fraction of the
LBGs correspond to subhalos (i.e., satellites).  This fraction is
sufficiently high for a significant deviation of the LBG correlation
function from the power law at small scales.

\end{itemize}

It is quite remarkable that, after vast improvements in observational
data over the past decade, the simple picture for the formation of
galaxies within virialized dark matter halos \citep{White78,Fall80},
whereby galaxy properties such as luminosity are tightly coupled to
the properties of their host halos, provides a good description of the
observed clustering trends of galaxies over nearly the entire age of
the Universe, on both large and small scales.

As we discussed above, our $L-${\Vmax} model has natural implications
for the Tully-Fisher relation and dynamical mass-to-light ratio, and
the evolution in these quantities. A fruitful followup to the study
presented here would be to compare the predictions of this model to
local and to (forthcoming) DEEP2 measurements of the relation.  The
comparison of model predictions to observed galaxy-galaxy and
galaxy-mass clustering can also constrain the amount of scatter in the
$L-${\Vmax} relation for bright galaxies, and readily predicts the
conditional luminosity function of galaxies (luminosity function as a
function of distinct halo mass), and its evolution.  Our luminosity
assignment scheme can also be extended by separately modeling central
and satellite galaxies, which appear to be two distinct populations
with different origins and observable properties
\citep{Vale04,Cooray05a,Cooray05b,Weinmann05,Hansen05}.

\acknowledgments The simulations were run on the Columbia machine at
NASA Ames and on the Seaborg machine at NERSC (Project PI: Joel
Primack). First and foremost, we would like to thank Anatoly Klypin
for running these simulations and making them available to us. We are
also indebted to Brandon Allgood for providing the merger trees that
were used to measure \Vin.  We thank Kurt Adelberger, Alison Coil,
Mauro Giavalisco, Kyoungsoo Lee, and Masami Ouchi for providing access
to and offering help with the interpretation of their data; and
especially thank Alison Coil and the DEEP team for making their
clustering data available to us in advance of publication. We
additionally thank Andrew Zentner and Jeremy Tinker for useful
discussions about the halo model and Oleg Gnedin for discussions about
GOODS clustering analysis and useful comments on a draft of this
manuscript.  RHW is supported by NASA through Hubble Fellowship grant
HST-HF-01168.01-A awarded by the Space Telescope Science
Institute. AVK is supported by the National Science Foundation (NSF)
under grants No.  AST-0206216, AST-0239759 and AST-0507666, and by
NASA through grant NAG5-13274.  This research was carried out at the
University of Chicago, Kavli Institute for Cosmological Physics and
was supported in part by the grant NSF PHY-0114422. KICP is an NSF
Physics Frontier Center.  This work made extensive use of the NASA
Astrophysics Data System and of the {\tt astro-ph} preprint archive at
{\tt arXiv.org}.

\end{document}